\documentclass[a4paper,11pt]{article}

\usepackage{blindtext,titlefoot}

\usepackage{amsfonts,amsthm,amssymb,amsmath}
\usepackage[title,titletoc,toc]{appendix}
\usepackage[utf8]{inputenc}
\usepackage[affil-it]{authblk}
\usepackage{xcolor}
\usepackage{tikz}
\newcommand{\RN}[1]{%
  \textup{\uppercase\expandafter{\romannumeral#1}}%
}
\newcommand\COMP{\hbox{C\kern -.58em {\raise .54ex \hbox{$\scriptscriptstyle |$}}
\kern-.55em {\raise .53ex \hbox{$\scriptscriptstyle |$}} }}
\newcommand\NN{\hbox{I\kern-.2em\hbox{N}}}
\newcommand\RR{\hbox{I\kern-.2em\hbox{R}}}
\newcommand\sRR{{\it \hbox{I\kern-.2em\hbox{R}}}}
\newcommand\QQ{\hbox{I\kern-.53em\hbox{Q}}}
\newcommand\PP{\hbox{I\kern-.53em\hbox{P}}}
\newcommand\EE{\hbox{I\kern-.53em\hbox{E}}}
\newcommand\ZZ{{{\rm Z}\kern-.28em{\rm Z}}}
\newcommand\be{\begin{equation}}
\newcommand\ee{\end{equation}}
\usepackage{scalerel,stackengine}
\stackMath
\newcommand\reallywidehat[1]{%
\savestack{\tmpbox}{\stretchto{%
  \scaleto{%
    \scalerel*[\widthof{\ensuremath{#1}}]{\kern.1pt\mathchar"0362\kern.1pt}%
    {\rule{0ex}{\textheight}}
  }{\textheight}%
}{2.4ex}}%
\stackon[-6.9pt]{#1}{\tmpbox}%
}

\newtheorem{theorem}{Theorem}[section]

\newtheorem{proposition}[theorem]{Proposition}
\newtheorem{remark}[theorem]{Remark}

\newtheorem{lemma}[theorem]{Lemma}

\newtheorem{corollary}[theorem]{Corollary}
\newtheorem{definition}[theorem]{Definition}

\makeatletter
\newcommand*\bigcdot{\mathpalette\bigcdot@{.5}}
\newcommand*\bigcdot@[2]{\mathbin{\vcenter{\hbox{\scalebox{#2}{$\m@th#1\bullet$}}}}}
\makeatother
\newcommand{\is}{\bigcdot }

\def \id{1\!\!1}

\def \Lbrack {[\![}
\def \Rbrack {]\!]}

\numberwithin{equation}{section}

\DeclareMathOperator*{\loc}{loc}

\begin{document}

\title{Representation for martingales living after a random time with applications}

\author{Tahir Choulli \\ 
Mathematical and Statistical Sciences Dept.\\
University of Alberta, Edmonton, Canada
\and 
Ferdoos Alharbi\\ 
Mathematical and Statistical Sciences Dept.\\
University of Alberta, Edmonton, Canada\\
and\\
Saudi Electronic University, Saudi Arabia

}

\maketitle\unmarkedfntext{
This research  is supported by NSERC (through grant NSERC RGPIN04987).\\
Address correspondence to Tahir Choulli, Department of Mathematical and Statistical Sciences, University of Alberta, 632 Central Academic Building,
Edmonton, Canada; e-mail: tchoulli@ualberta.ca}

\begin{abstract}Our financial setting consists of a market model with two flows of information. The smallest flow $\mathbb{F}$ is the ``public" flow of information which is available to all agents, while the larger flow $\mathbb{G}$ has additional information about the occurrence of a random time $\tau$. This random time can model the default time in  credit risk or death time in life insurance. Hence the filtration $\mathbb{G}$ is the progressive enlargement of $\mathbb{F}$ with $\tau$. In this framework, when $\tau$ is a finite honest time, we describe explicitly how $\mathbb G$-local martingales can be represented in terms of $\mathbb{F}$-local martingales and parameters of $\tau$. This representation complements Choulli, Daveloose and Vanmaele \cite{ChoulliDavelooseVanmaele} to the case when martingales live ``after $\tau$". Under some mild assumptions on the pair $(\mathbb{F}, \tau)$, we fully elaborate the application of these results to the explicit parametrization of all deflators under $\mathbb{G}$. The results are illustrated in the case of a jump-diffusion model and a discrete-time market model.
\end{abstract}


\section{Introduction} 
This paper considers a class of informational markets, which is defined by the pair $(\mathbb{F},\tau)$. Herein, $\mathbb{F}$ models the ``public" information that is available to all agents over time, while $\tau$ is a random time that might not be seen through the flow $\mathbb{F}$ when it occurs. This random time represents a default time of a firm in credit risk theory, a death time of an agent in life insurance where the mortality and longevity risks pose serious challenges, or the occurrence time of an event that might impact the market somehow. For detailed discussion about the relationship between our current framework with the credit risk literature, we refer the reader to  Choulli et al. \cite{ChoulliDavelooseVanmaele}. As random times can be seen before their occurrence, the flow of the agents who can see $\tau$ happening results from the progressive enlargement of $\mathbb{F}$ with $\tau$, and which will be denoted by $\mathbb{G}$ throughout the rest of the paper. \\

In this setting, our first principal objective lies in quantifying the various risks  induced by $\tau$ and its ``correlation" with $\mathbb{F}$. Mathematically, usually a risk is represented by a random variable which can be seen as a terminal value of a martingale (the dynamic version of this risk). Thus, our objective boils down to elaborating the following representation for any $\mathbb{G}$-martingale $M^{\mathbb{G}}$ 
\begin{equation}\label{Martingale Representation4Intro}
M^{\mathbb{G}}=M^{(\mbox{pf})}+M^{(\mbox{pd},1)}+...+M^{(\mbox{pd},k)}+M^{(\mbox{cr},1)}+...+M^{(\mbox{cr},l)}.
\end{equation}
All terms in the right-hand side are $\mathbb{G}$-local martingales, where $M^{(\mbox{pf})}$ represents the pure financial risk,  $M^{(\mbox{pd},i)}$ $i=1,...,k$ are the pure default/mortality risks, and   $M^{(\mbox{cr},j)}$ $j=1,...,l$ are the correlation risks. The representation (\ref{Martingale Representation4Intro}) appeared first in Az\'ema et al. \cite{Azema} when $\mathbb{F}$ is a Brownian filtration and $\tau$ avoid $\mathbb{F}$-stopping times and is the end of an $\mathbb{F}$-predictable set. Then Blanchet-Scalliet and Jeanblanc (2004) focused on a restricted subset of $\mathbb{G}$-martingales stopped at $\tau$ and extends slightly the representation under a set of assumptions on the pair $(\mathbb{F},\tau)$. Recently, the set of $\mathbb{G}$-martingales stopped at $\tau$ was fully elaborated under no assumption in  \cite{ChoulliDavelooseVanmaele}. The applications of this representation in credit risk can be found in \cite{BelangerShreveWong,BieleckiRutkowski,BlanchetJeanblanc} (see also \cite{ChoulliDavelooseVanmaele} for more related literature), while its application in arbitrage can be found in \cite{ChoulliYansori}. Up to our knowledge, for $\mathbb{G}$-martingales living after $\tau$ and general pair $(\mathbb{F},\tau)$, the martingales representation formula (\ref{Martingale Representation4Intro}) remains an open question.\\

This paper assumes that $\tau$ is an honest time, and elaborates the formula (\ref{Martingale Representation4Intro}) for $\mathbb{G}$-martingales living after $\tau$, and hence it complements the study considered in \cite{ChoulliDavelooseVanmaele}. Then, by combining these obtained results with \cite{ChoulliDavelooseVanmaele}, we derive the exact form of (\ref{Martingale Representation4Intro}) for honest times. This extends \cite{Azema} to a more general setting for the pair $(\mathbb{F},\tau)$.\\

Our second objective in this paper resides in deriving direct applications of the representation (\ref{Martingale Representation4Intro})  to the explicit description of
 deflators for the models $(S-S^{\tau}, \mathbb{G})$  in terms of the deflators of the initial model $(S,\mathbb{F})$. Here $S$ is the discounted price processes of $d$-risky assets, which is mathematically an $\mathbb{F}$-semimartinagle. This complements \cite{ChoulliYansori}, which focuses on market models stopped at $\tau$. Therefore, again, we combined our obtained results on deflators with \cite{ChoulliYansori} to describe the set of all deflators for the model $(S,\mathbb{G})$ afterwards. \\
 
 This paper  contains four sections including the current introduction section. The second section presents the mathematical model, its parametrization and some preliminaries that will be used throughout the paper. The third section addresses our first goal and gives results about the representation (\ref{Martingale Representation4Intro}). The fourth section deals with the second main objective of deflators descriptions. The paper has some appendices where we recall some known results for the sake of having a self-contained paper, and where we relegate some technical proofs. 
\section{The mathematical framework and preliminaries}\label{Section2} Throughout the paper, we consider given a complete probability space  $\left(\Omega, {\cal{G}},P\right)$.  
For any filtration  $\mathbb H\in \{\mathbb F,\mathbb G\}$ defined on this space, we denote ${\cal A}(\mathbb H)$ (respectively ${\cal M}(\mathbb H)$) the set
of $\mathbb H$-adapted processes with $\mathbb H$-integrable variation (respectively that are $\mathbb H$-uniformly integrable martingale), and the total variation process of a process $X$ -when it exists- will be denoted throughout the paper by Var$(X)$.
For any process $X$, we denote by $^{o,\mathbb H}X$  (respectively $^{p,\mathbb H}X$)  the
$\mathbb H$-optional (respectively $\mathbb H$-predictable) projection of $X$. For an increasing process $V$, we denote $V^{o,\mathbb H}$ (respectively $V^{p,\mathbb H}$) its dual $\mathbb H$-optional (respectively dual $\mathbb H$-predictable) projection. For a filtration $\mathbb H$, ${\cal O}(\mathbb H)$, ${\cal P}(\mathbb H)$ and  $\mbox{Prog}(\mathbb H)$ represent the $\mathbb H$-optional, the $\mathbb H$-predictable and the $\mathbb H$-progressive $\sigma$-fields  respectively on $\Omega\times[0,+\infty[$. For an $\mathbb H$-semimartingale $X$, we denote by $L(X,\mathbb H)$ the set of all $X$-integrable processes in Ito's sense, and for $H\in L(X,\mathbb H)$, the resulting integral is one dimensional $\mathbb H$-semimartingale denoted by $H\is X:=\int_0^{\cdot} H_udX_u$. For $M \in {\cal{M}}_{loc}(\mathbb{H})$, we denote $L^1_{loc}(M,\mathbb{H})$ the subset of  $L(M,\mathbb H)$ such that the resulting integral is an $\mathbb{H}$-local martingale. Throughout the paper, for any process $X$ and any random time $\theta$, we denote by $X^{\theta}$ the stopped process at $\theta$ given by $X^{\theta}_t:=X_{t\wedge\theta}$, $t\geq 0$. If ${\cal C}(\mathbb H)$ 
is a set of $\mathbb H$-adapted processes,
then ${\cal C}_{\loc}(\mathbb H)$ --except when it is stated otherwise-- is the set of processes, $X$,
for which there exists a sequence of $\mathbb H$-stopping times,
$(T_n)_{n\geq 1}$, that increases to infinity and $X^{T_n}$ belongs to ${\cal C}(\mathbb H)$, for each $n\geq 1$. For any $\mathbb H$-semimartingale, $L$, the Dol\'eans-Dade stochastic exponential denoted by ${\cal E}(L)$, is the unique solution to the SDE: $dX = X_{-} dL,$ $ X_0= 1,$ given by 
\begin{equation}\label{S-exponential}
{\cal E}_t (L) = \exp \big ( L_t-L_0 - {1 \over 2} {\langle L^c \rangle}_{t} \big ) \prod_{0 <  s \leq t} \big( 1 + \Delta L_s \big) e^{-\Delta L_s}.
\end{equation}

Our mathematical model starts with a stochastic basis $\left(\Omega, {\cal{G}},\mathbb{F}=({\cal{F}}_t)_{t\geq 0},P\right)$, where $\mathbb{F}$ is a filtration satisfying the usual hypothesis (i.e., right continuity and completeness) and ${\cal{F}}_{\infty}\subset{\cal{G}}$. Financially speaking, the filtration $\mathbb{F}$ represents the flow of ``public information" through time. Besides this initial model, we consider a random time $\tau$, i.e., a $[0,+\infty]$-valued $\cal{G}$-measurable random variable. To this random time, we associate the process $D$ and the filtration $\mathbb{G}$ given by 
\begin{equation}\label{ProcessDandG}
D := I_{\Lbrack\tau,+\infty\Lbrack},\ \ \ \mathbb G:=({\cal G}_t)_{t\geq 0},\ \ \ {\cal G}_t=
\cap_{s>0}\left({\cal F}_{s+t}\vee\sigma\left(D_{u},\ u\leq s+t\right)\right).\end{equation}
The agent endowed with $\mathbb F$ can only get information about $\tau$ via the {\it survival probabilities} $G$ and $\widetilde G$, known in the literature as Az\'ema supermartingales, and are given by
\begin{eqnarray}\label{GGtilde}
G_t :=^{o,\mathbb F}(I_{\Lbrack0,\tau\Lbrack})_t= P(\tau > t | {\cal F}_t) \ \mbox{ and } \ \widetilde{G}_t :=^{o,\mathbb F}(I_{\Lbrack0,\tau\Rbrack})_t= P(\tau \ge t | {\cal F}_t).\end{eqnarray}
Throughout the paper, besides the pair $(G,{\widetilde{G}})$ that parametrizes $\tau$, the following process
\begin{equation} \label{mprocess}
m := G + D^{o,\mathbb F},
\end{equation}
plays a central role in our analysis and it is a BMO $\mathbb F$-martingale. For more details about this and other related results, we refer the reader to  \cite[paragraph 74, Chapitre XX]{dellacheriemeyer92}.\\

In this paper, we focus on the class of honest times introduced in \cite{barlow}, and which we recall below.
\begin{definition}\label{honesttime}
 A random time $\sigma$ is called an ${\mathbb F}$-honest time if, for any $t$, there exists an ${\cal{F}}_t$-measurable random variable $\sigma_t$ such that $\sigma{I}_{\{\sigma<t\}} = \sigma_t{I}_{\{\sigma<t\}}.$
\end{definition}

 The following theorem introduces two different classes of $\mathbb{G}$-martingales, which are very useful in our analysis. The  first class is intimately related to an operator which transforms $\mathbb{F}$-martingales into $\mathbb{G}$-martingales, and this operator appeared naturally in the representation of $\mathbb{G}$-martingales. The second class consists of the $\mathbb{G}$-local martingale part  of $M-M^{\tau}$, when $M$ spans the set of $\mathbb F$-local martingales.
 \begin{theorem} \label{OptionalDecompoTheorem}  Suppose $\tau$ is an honest time. Then the following assertions hold.\\
 {\rm{(a)}} For any $\mathbb F$-local martingale $M$, the process
 \begin{eqnarray}\label{T(M)afterTau}
{\cal T}^{(a)}(M):=I_{\Rbrack\tau,+\infty\Lbrack}\is{M}+{{I_{\Rbrack\tau,+\infty\Lbrack}}\over{1- G}}\is [m,M]+{{I_{\Rbrack\tau,+\infty\Lbrack}}\over{1-G_{-}}}\is \left(\sum \Delta M(1-G_{-}) I_{\{\widetilde G=1>G_{-}\}}\right)^{p,\mathbb F}
 \end{eqnarray}
 is a $\mathbb G$-local martingale.\\
 {\rm{(b)}} For any $M \in {\mathcal M}_{loc} (\mathbb F)$, the process 
\begin{equation}\label{Glocalmartingaleaftertau}
  \widehat{M}^{(a)}:= I_{\Rbrack \tau, +\infty \Lbrack} \is M + (1-G_{-})^{-1} I_{\Rbrack \tau, +\infty \Lbrack} \is \langle m, M\rangle^{\mathbb{F}}\quad\mbox{is a $\mathbb G$-local martingale.}
\end{equation}
 \end{theorem}
 
The proof of assertion (a) can be found in \cite[Proposition 4.3]{ChoulliDeng}, while assertion (b) is given in \cite[Lemma 2.6]{aksamitetal18} (or see \cite[Th\'eor\`eme 5.10]{jeulin80} and \cite[XX.79]{dellacheriemeyer92} for this assertion and related results). The superscript in the operator ${\cal T}^{(a)}$ refers to the case of ``after $\tau$", while in ${\cal{T}}^{(b)}$ --which will be defined later in Theorem \ref{Toperator}-- refers to the case of ``before or at $\tau$". 
 
 \section{Martingale representation theorems}\label{SecRepresentation1time}
 This section complements the work of Choulli et al. \cite{ChoulliDavelooseVanmaele} and parametrizes fully and explicitly the $\mathbb{G}$-local martingales that live after $\tau$, which we assume being an honest time. This section is divided into three subsections. The first subsection presents our main representation results  for $\mathbb{G}$-local martingales that live after $\tau$, and illustrates those results on two particular cases. The second subsection combines the first subsection with Choulli et al. \cite{ChoulliDavelooseVanmaele} in order to derive a full and complete representation of general $\mathbb{G}$-martingales, while the last subsection proves the main results of the first subsection. 
 \subsection{The case of martingales living after $\tau$}\label{Subsection4RepresentationAfterTau}
  This subsection extends the main result of \cite{ChoulliDavelooseVanmaele} to $\mathbb G$-local martingales which live on the stochastic interval $\Rbrack\tau,+\infty\Lbrack$. It shows how to represent $\mathbb G$-local martingales on $\Rbrack\tau,+\infty\Lbrack$ using $\mathbb F$-local martingales.

 \begin{theorem}\label{TheoRepresentation00}  Let $m$ be defined in  (\ref{mprocess}), and suppose $\tau$ is an honest time. Consider a RCLL and $\mathbb{G}$-adapted process $M^{\mathbb{G}}$. Then the following assertions hold.\\
   {\rm{(a)}}  $I_{\Rbrack\tau,+\infty\Lbrack}\is{M}^{\mathbb G}$ is a square integrable $\mathbb G$-martingale (i.e.   $I_{\Rbrack\tau,+\infty\Lbrack}\is{M}^{\mathbb G}\in {\cal{M}}^2(\mathbb{G})$) if and only if there exists a unique square integrable $\mathbb{F}$-martingale $M$ satisfying 
 \begin{equation}\label{Representation200}
  I_{\{G_{-}=1\}}\is M=0,\quad  \Delta{M}{I}_{\{\widetilde G=1\}}=0,\end{equation}
  \begin{equation}\label{Representation200bis} 
\displaystyle{E}\left[\int_0^{\infty}I_{\{\widetilde{G}_s<1\}}(1-\widetilde{G}_s)^{-1}d[M,M]_s\right]<\infty,\end{equation}
and
\begin{equation}\label{Representation201}
(1-G_{-})I_{\Rbrack\tau,+\infty\Lbrack}\is{M}^{\mathbb G}={\cal T}^{(a)}(M).
\end{equation}
 {\rm{(b)}}  $I_{\Rbrack\tau,+\infty\Lbrack}\is{M}^{\mathbb G}$ is a $\mathbb G$-local martingale with integrable variation (i.e.   $I_{\Rbrack\tau,+\infty\Lbrack}\is{M}^{\mathbb G}\in {\cal{M}}_{loc}(\mathbb{G})\cap{\cal{A}}(\mathbb{G})$) if and only if there exists a unique $M\in{\cal{M}}_{loc}(\mathbb{F})$ with integrable variation satisfying (\ref{Representation200}) and (\ref{Representation201}).
\end{theorem}
The proof of Theorem \ref{TheoRepresentation00} is relegated to Subsection \ref{Subsection4proofs} for the sake of easy exposition. Theorem \ref{TheoRepresentation00} gives martingale parameterizations which are unique, explicit and complete no matter what is the model for the honest time $\tau$. It is important to mention that, as stated in \cite[Proposition (5.1), Chapter V]{jeulin80} , $\tau$ being an honest time is in general weaker than $\tau$ being the {\it end of an $\mathbb{F}$-optional random set}. In fact, these two concepts (honest time and being the end of an optional random set) coincide for {\it finite} random times only (i.e. when $\tau<+\infty$ $P$-a.s.). On the one hand, without this finiteness property of $\tau$, the localization property might not remain valid when passing from $\mathbb{G}$ to $\mathbb{F}$. It is exactly this point that forced us to consider the two classes of square integrable martingales and martingales with integrable variations.  On the other hand, for the case of finite honest times, we can do much better and elaborate the localization version of Theorem \ref{TheoRepresentation00} (i.e., we can drop the assumption of uniform integrability on $M^{\mathbb{G}}$). This is the aim of the  next representation theorem. To this end, we consider the following subspace of $\mathbb{F}$-local martingales, which {\it naturally} appeared in our analysis and it is very intrinsic to the pair $(\mathbb{F},\tau)$,
 \begin{equation}\label{M(F,Tau)}
{\mathbb{M}}{(\mathbb{F},\tau)}:=\left\{M\in {\cal{M}}_{0,loc}(\mathbb{F})\ :\quad E\left[\int_0^{\infty}{{I_{\{\widetilde{G}_s<1\}}}\over{1-\widetilde{G}_s+\vert\Delta{M}_s\vert}} d[M,M]_s\right]<+\infty\right\}.
 \end{equation}
Thus, in virtue of the localization defined in Section \ref{Section2}, we have 
$$M\in {\mathbb{M}}_{loc}(\mathbb{F},\tau)\quad \mbox{if and only if}\quad I_{\{\widetilde{G}_s<1\}}\left(1-\widetilde{G}_s+\vert\Delta{M}_s\vert\right)^{-1} \is [M,M]\in {\cal{A}}^+_{loc}(\mathbb{F}).$$
To explain how natural this space appears in our coming results and analysis, it is worth comparing it to the space of $\mathbb{F}$-martingales of Theorem \ref{TheoRepresentation00}-(a). To this end, we consider the case when all $\mathbb{F}$-martingales are continuous, and easily conclude that in this setting the two spaces coincide.  In general,  it is clear that any $\mathbb{F}$-local martingale $M$ satisfying (\ref{Representation200bis}) belongs to ${\mathbb{M}}{(\mathbb{F},\tau)}$. Below, we summarize other properties of ${\mathbb{M}}{(\mathbb{F},\tau)}$, which are interesting in themselves.
\begin{remark}\label{Remark4M(F,Tau)}{\rm{(a)}} If $M\in {\cal{M}}_{0,loc}(\mathbb{F})$ with integrable variation, then it belongs to  ${\mathbb{M}}{(\mathbb{F},\tau)}$, or equivalently $ {\cal{M}}_{0,loc}(\mathbb{F})\cap{\cal{A}}(\mathbb{F})\subset{\mathbb{M}}{(\mathbb{F},\tau)} $. This claim follows from combining the fact that ${\rm{Var}}(M)=\sum\vert\Delta{M}\vert$ (see \cite[Lemma 2.61, Chapter 2]{Jacod}) and the following inequalities
\begin{equation*}
{{I_{\{\widetilde{G}<1\}}}\over{1-\widetilde{G}+\vert\Delta{M}\vert}} \is[M,M]= {{\vert\Delta{M}\vert}\over{1-\widetilde{G}+\vert\Delta{M}\vert}}I_{\{\widetilde{G}<1\}}\is {\rm{Var}}(M)\leq {\rm{Var}}(M).
\end{equation*}
{\rm{(b)}} Suppose $M$ is a continuous $\mathbb{F}$-local martingale. Then  $M\in{\mathbb{M}}{(\mathbb{F},\tau)}$ if and only if there exists a continuous and square integrable $\mathbb{F}$-martingale $N$ satisfying $M=\sqrt{1-G_{-}}\is N$. As a result, if ${\cal{M}}_{loc}^c(\mathbb{F})$ denotes the set of all $\mathbb{F}$-local martingale that are continuous, then we get 
$$\left\{\sqrt{1-G_{-}}\is N\ :\ N\ \mbox{is a continuous and square integrable $\mathbb{F}$-martingale}\right\}={\mathbb{M}}{(\mathbb{F},\tau)}\cap{\cal{M}}_{loc}^c(\mathbb{F}).$$
\end{remark}
Now, we state our second representation theorem for general {\it finite honest times}.
 \begin{theorem}\label{TheoRepresentation0}  Let $m$ be defined in  (\ref{mprocess}), and suppose
  \begin{equation}\label{MainAssumption4TauBis}
 \tau\quad\mbox{is a finite honest time}\quad ({\rm{i.e.}}\ \tau<+\infty\quad P\mbox{-a.s.}).\end{equation} 
Consider a RCLL and $\mathbb{G}$-adapted process $M^{\mathbb{G}}$. Then the following assertions are equivalent.\\
   {\rm{(a)}}  $I_{\Rbrack\tau,+\infty\Lbrack}\is{M}^{\mathbb G}$ is a $\mathbb G$-local martingale (i.e.   $I_{\Rbrack\tau,+\infty\Lbrack}\is{M}^{\mathbb G}\in {\cal{M}}_{loc}(\mathbb{G})$).\\
 {\rm{(b)}} There exists an $\mathbb F$-local martingale $M^{\mathbb F}$ satisfying 
 \begin{equation}\label{RepresentationConditionsBis}
M^{\mathbb F}I_{\{\widetilde G=1\}}=0,\  {{M^{\mathbb{F}}_{-}I_{\{G_{-}<1\}}}\over{1-G_{-}}}\ \mbox{is $\mathbb{F}$-locally bounded},\quad\displaystyle{I}_{\{G_{-}<1\}}\is{M}^{\mathbb{F}}+{{M^{\mathbb{F}}_{-}I_{\{G_{-}<1\}}}\over{1-G_{-}}}\is m\in {\mathbb{M}}_{loc}{(\mathbb{F},\tau)},
\end{equation}
and
\begin{equation}\label{Representation22Bis}
(1-G_{-})I_{\Rbrack\tau,+\infty\Lbrack}\is{M}^{\mathbb G}= {\cal T}^{(a)}(M^{\mathbb F})
+ {{M^{\mathbb F}_{-}}\over{1-G_{-}}} I_{\Rbrack\tau,+\infty\Lbrack} \is{\cal T}^{(a)}(m).
\end{equation}
{\rm{(c)}} There exists a unique $M\in {\mathbb{M}}_{loc}{(\mathbb{F},\tau)}$ such that
 \begin{equation}\label{Representation222Bis}
  I_{\{G_{-}=1\}}\is M=0,\quad  \Delta{M}{I}_{\{\widetilde G=1\}}=0\quad\mbox{and}\quad  (1-G_{-})I_{\Rbrack\tau,+\infty\Lbrack}\is{M}^{\mathbb G}={\cal T}^{(a)}(M).
\end{equation}
\end{theorem}
The proof of this theorem is relegated to Subsection \ref{Subsection4proofs}. It is worth mentioning that our Theorem \ref{TheoRepresentation0} can be extended (without much difficulties) to the set of $\mathbb{G}$-local martingale that are locally $p$-integrable (for any $p\in (1,+\infty)$). Indeed, if we denote ${\cal{M}}_{loc}^p(\mathbb{G})$  such a set, and we consider 
$${\mathbb{M}}^{(p)}_{loc}{(\mathbb{F},\tau)}:=\left\{M\in {\cal{M}}_{0,loc}(\mathbb{F})\ :\quad \left({{I_{\{\widetilde{G}<1\}}}\over{1-\widetilde{G}+\vert\Delta{M}\vert}} \is [M,M]\right)^p\in{\cal{A}}^+_{loc}(\mathbb{F})\right\},$$
then one can state the similar theorem by substituting ${\cal{M}}_{loc}^p(\mathbb{G})$ and ${\mathbb{M}}^{(p)}_{loc}{(\mathbb{F},\tau)}$ for  ${\cal{M}}_{loc}(\mathbb{G})$ and ${\mathbb{M}}_{loc}{(\mathbb{F},\tau)}$ respectively. This case of (locally) $p$-integrable local martingales is mainly motivated by the application in arbitrage theory, and precisely in describing martingale densities/deflators  (see Definition \ref{DeflatorDefinition} for deflators) which are (locally) $p$-integrable. For more on these and their applications, we refer the reader to \cite{ChoulliStricker99, Choulli2007,DelbaenSchachermayer96b,Stricker90} and the references therein. \\
Theorem \ref{TheoRepresentation0} gives deep insight on the structures of the representation (see assertion (b)), and how $\tau$ affects the martingale structures through both the process $m$ and the operator ${\cal{T}}^{(a)}$. By using this structure, we lost the uniqueness of the parametrization which holds on the set $(G_{\tau}<1)$ only. \\
From the perspective of applications in arbitrage theory, as it is our concerns herein, it was proven in the literature that an arbitrary finite honest time $\tau$ might trigger arbitrage opportunities in the part after $\tau$ no matter how nice the initial model is. To address this issue, \cite{aksamitetal18} introduced a subclass of honest times, which is rich enough and has the possibility of preserving the non-arbitrage condition after $\tau$ for a large class of initial models. Thus, the aim of the next representation theorem lies in showing that when we restrict to this subclass of honest times, the parametrization of $\mathbb{G}$-local martingales requires the full space of $\mathbb{F}$-local martingales instead.
 \begin{theorem}\label{TheoRepresentation1}   Consider a process $M^{\mathbb{G}}$, and suppose that (\ref{MainAssumption4TauBis}) holds and 
 \begin{equation}\label{MainAssumption4Tau}
 G_{\tau}<1\quad P\mbox{-a.s.}.\end{equation}
  Then the following assertions are equivalent.\\
 {\rm{(a)}} $I_{\Rbrack\tau,+\infty\Lbrack}\is{M}^{\mathbb G}\in {\cal M}_{loc}(\mathbb{G})$.\\
 {\rm{(b)}}  There exists a unique $\widetilde{M}^{\mathbb F}\in {\cal M}_{0,loc}(\mathbb{F})$ satisfying $ \widetilde{M}^{\mathbb F}I_{\{\widetilde G=1\}}=0$ and 
\begin{equation}\label{Representation22}
I_{\Rbrack\tau,+\infty\Lbrack}\is{M}^{\mathbb G}=\displaystyle{{I_{\Rbrack\tau,+\infty\Lbrack}}\over{1-G_-}}  \is {\cal T}^{(a)}(\widetilde{M}^{\mathbb F})
+ {{\widetilde{M}^{\mathbb F}_{-}}\over{(1-G_{-})^2}} I_{\Rbrack\tau,+\infty\Lbrack} \is{\cal T}^{(a)}(m).
\end{equation}
{\rm{(c)}} There exists a unique $M\in{\cal M}_{0,loc}(\mathbb{F})$ such that
 \begin{equation}\label{Representation222}
  I_{\{G_{-}=1\}}\is M=0,\quad  {I}_{\{\widetilde G=1\}}\Delta{M}=0,\quad   I_{\Rbrack\tau,+\infty\Lbrack}\is{M}^{\mathbb G}={\cal T}^{(a)}(M).
\end{equation}
\end{theorem}
The rest of this subsection illustrate Theorems \ref{TheoRepresentation00} , \ref{TheoRepresentation0}  and \ref{TheoRepresentation1}  to the case when the initial model is a jump-diffusion model  or  a discrete-time market model.
 \begin{corollary} Suppose $\mathbb F$ is the augmented filtration of the filtration generated by $(W,N)$, where $W$ is a standard  Brownian motion, $N$ is the Poisson process with intensity one, and $N_t^{\mathbb F}:=N_t-t$. Consider the pair $(\phi^{(m)},\psi^{(m)})\in{L}^1_{loc}(W,\mathbb F)\times L^1_{loc}(N^{\mathbb F},\mathbb F)$ such that 
 \begin{equation}\label{Representation4m}
 m=m_0+ \phi^{(m)}\is W+\psi^{(m)}\is N^{\mathbb{F}}.\end{equation}
 Let $M^{\mathbb G}$ be a RCLL process. Then the following assertions hold. \\
 {\rm{(a)}}  $M^{\mathbb G}\in {\cal M}_{0}^2(\mathbb{G})$ if and only if there exists a unique pair of predictable processes $(\phi,\psi)$ satisfying
 \begin{equation}\label{Condition4JumpDiffusion0}
 (\phi,\psi)I_{\{G_{-}=1\}}\equiv(0,0),\quad \psi I_{\{\psi^{(m)}=1-G_{-}\}}\equiv 0,\end{equation}
  \begin{equation}\label{Condition4JumpDiffusion1}
E\left[\int_0^{\infty}{{\phi_s^2{I}_{\{G_{s}<1\}}}\over{1-G_{s}}}ds +\int_0^{\infty} {{\psi_s^2}\over{1-G_{s}-\psi^{(m)}_s}}I_{\{G_{s}+\psi^{(m)}_s<1\}}ds \right]<\infty,\end{equation}
  and 
  \begin{equation}\label{Representation4JumpDiffusion}
  M^{\mathbb G}-\left(M^{\mathbb G}\right)^{\tau}={{\phi}\over{1-G_{-}}}I_{\{G_{-}<1\}}\is{\cal T}^{(a)}(W)+{{\psi}\over{1-G_{-}}}I_{\{G_{-}<1\}}\is{\cal T}^{(a)}(N^{\mathbb F}).\end{equation}
 {\rm{(b)}} Suppose that (\ref{MainAssumption4TauBis}) holds. Then for any $M^{\mathbb G}\in {\cal M}_{0,loc}(\mathbb{G})$, there exists a unique pair of predictable processes $(\phi,\psi)$ satisfying (\ref{Condition4JumpDiffusion0}), (\ref{Representation4JumpDiffusion}) and 
  \begin{equation}\label{Condition4JumpDiffusion1.2}
 V_t:= \int_0^t\left({{\phi_s^2{I}_{\{G_{s}<1\}}}\over{1-G_{s}}}+{{\psi_s^2}\over{1-G_{s}-\psi^{(m)}_s}}I_{\{G_{s}+\psi^{(m)}_s<1\}}\right)ds\in {\cal{A}}^+_{loc}(\mathbb{F}).\end{equation}
  {\rm{(c)}} Suppose that both (\ref{MainAssumption4TauBis}) and (\ref{MainAssumption4Tau}) hold.  Then for any $M^{\mathbb G}\in {\cal M}_{0,loc}(\mathbb{G})$, there exists a unique pair $(\phi_1,\psi_1)$ of $\mathbb{F}$-predictable processes such that $\int_0^{\cdot}((\phi_1(s))^2+(\psi_1(s))^2)ds$ is locally integrable and 
 \begin{equation}\label{Representation4JumpDiffusion1}
  M^{\mathbb G}-\left(M^{\mathbb G}\right)^{\tau}=\phi_1\is{\cal T}^{(a)}(W)+{\psi}_1\is{\cal T}^{(a)}(N^{\mathbb F}).\end{equation}
 \end{corollary}
 
\begin{proof} Assertion (a) is  a direct consequence of   Theorem \ref{TheoRepresentation00}-(a) and the fact that for any $\mathbb F$-local martingale, $M$, there exists a unique pair
 $(\varphi_1,\varphi_2)$ that is $(W,N^{\mathbb{F}})$-integrable and $M=M_0+\varphi_1\is W+\varphi_2\is N^{\mathbb F}$.\\
 Again by combining this fact with Theorem \ref{TheoRepresentation0}  (respectively with Theorem \ref{TheoRepresentation1}), assertion (b) (respectively assertion (c)) follows. This proves the corollary.\end{proof}
 
The second popular case for $(S,\mathbb F)$ is the discrete-time model, where we  suppose that on $(\Omega, {\cal F},P)$ the following assumptions hold.
\begin{eqnarray}\label{GandG*}
P(\tau\in\{0,1,..., T\})=1,\quad   \mathbb{F}:= ({\cal F}_n)_{n=0,1,..., T},\ {\cal G}_n={\cal F}_n\vee \sigma\left(\tau\wedge 1,...,\tau\wedge n\right),\end{eqnarray}
As a result, in this case, the pair $(G,\widetilde{G})$ that parametrizes $\tau$ in $\mathbb{F}$ takes the following forms
\begin{equation}\label{OpertaorTdiscrete}G_n=\displaystyle\sum_{k=n+1}^T P(\tau =k | {\cal F}_n),\quad n=0,...,T-1, \quad\mbox{and}\quad \widetilde{G}_n=\sum_{k=n}^T P(\tau =k | {\cal F}_n),\ n=0,...,T.\end{equation}
\begin{corollary} Consider the model given by (\ref{GandG*}), and suppose that $\tau$ is an honest time. Then for any $\mathbb{G}$-local martingale $M^{\mathbb{G}}$, there exists a unique $\mathbb{F}$-local martingale $M^{\mathbb{F}}$ satisfying $M^{\mathbb{F}}_0=0$,
\begin{equation}\label{Condition4MF}
\Delta M^{\mathbb{F}}_n I_{\{\widetilde{G}_n=1\}}:=(M^{\mathbb{F}}_n-M^{\mathbb{F}}_{n-1})I_{\{\widetilde{G}_n=1\}}\equiv 0,\quad P\mbox{-a.s.}\quad n=1,...,T,\end{equation}
and
\begin{equation}\label{Representation4MG-Discrete}
M^{\mathbb{G}}_n-M^{\mathbb{G}}_{n\wedge\tau}=\sum_{k=1}^n{{\Delta M^{\mathbb{F}}_k}\over{1-\widetilde{G}_k}}I_{\{\tau<k\}}= \sum_{k=1}^n{{\Delta{\cal T}^{(a)}_k(M^{\mathbb{F}})}\over{1-G_{k-1}}}I_{\{\tau<k\}},\quad n=0,...,T.
\end{equation}
\end{corollary}
\begin{proof} We start by proving the uniqueness. In fact, this boils down to proving that any $\mathbb F$-local martingale $M$ satisfying (\ref{Condition4MF}) and  
$${{\Delta M^{\mathbb{F}}_k}\over{1-\widetilde{G}_k}}I_{\{\tau<k\}}=0,\quad P\mbox{-a.s.},\quad k=1,...,T,$$
$M$ should be null. By taking the conditional expectation with respect to ${\cal F}_k$ on both sides of the above equality, we get for any $k=1,...,T$,
$$(\Delta M^{\mathbb{F}}_k)I_{\{\widetilde{G}_k<1\}}=0,\quad P\mbox{-a.s}.$$
Thus by combining this with (\ref{Condition4MF}) and $M^{\mathbb{F}}_0=0$, we get 
$$(\Delta M^{\mathbb{F}}_k)=(\Delta M^{\mathbb{F}}_k)I_{\{\widetilde{G}_k<1\}}+(\Delta M^{\mathbb{F}}_k)I_{\{\widetilde{G}_k=1\}}=0,\quad k=1,...,T,$$
and the uniqueness is proved. For the rest of this proof,  we consider the following processes
\begin{equation*}
\overline{M}^{\mathbb{G}}_k:=M^{\mathbb{G}}_k-M^{\mathbb{G}}_{k\wedge\tau},\quad{\rm{and}}\quad  \Delta{M}_k:=E\left[\Delta\overline{M}^{\mathbb{G}}_k\ \big|{\cal F}_k\right],\quad k=1,..., T,\quad M_0=E\left[\overline{M}^{\mathbb{G}}_0\ \Big|{\cal F}_0\right].
\end{equation*}
Then, it is easy to check that $\overline{M}^{\mathbb{G}}$ is a $\mathbb{G}$-local martingale, $M$ is an $\mathbb{F}$-local martingale, and 
\begin{equation}\label{MGproperty}
\Delta\overline{M}^{\mathbb{G}}_n=(\Delta\overline{M}^{\mathbb{G}}_n) I_{\{\tau<n\}},\quad P\mbox{-a.s.},\quad n=1,...,T.\end{equation}
Furthermore, in virtue of this latter equality and the fact that $(\tau<n)\subset(\widetilde{G}_n<1)\subset(G_{n-1}<1)$ $P$-a.s., we deduce that $(\Delta\overline{M}^{\mathbb{G}}_n)I_{\{\widetilde{G}_n=1\}}=0$ $P$-a.s. and hence we get 
$$
 (\Delta{M}_n)I_{\{\widetilde{G}_n=1\}}=E\left[(\Delta\overline{M}^{\mathbb{G}}_n)I_{\{\widetilde{G}_n=1\}}\ \big|{\cal F}_n\right]=0,\quad P\rm{-a.s.}.$$
 This proves the existence of an $\mathbb F$-local martingale $M$ satisfying (\ref{Condition4MF}). Now we prove that this $M$ satisfies (\ref{Representation4MG-Discrete}). To this end, we combine (\ref{MGproperty}) with \cite[Proposition (5.3)-(b)]{jeulin80}, and derive 
$$
{{ \Delta{M}_n}\over{1-\widetilde{G}_n}}I_{\{\tau<n\}}={{E\left[\Delta\overline{M}^{\mathbb{G}}_n I_{\{\tau<n\}}\ \big|{\cal F}_n\right]}\over{1-\widetilde{G}_n}}I_{\{\tau<n\}}=E\left[\Delta\overline{M}^{\mathbb{G}}_n I_{\{\tau<n\}}\ \big|{\cal G}_n\right]=\Delta\overline{M}^{\mathbb{G}}_n.$$
Thus, the first equality in (\ref{Representation4MG-Discrete}) follows immediately, while the second equality is a direct consequence of (\ref{Condition4MF}) combined with  
$$
{\cal{T}}^{(a)}_n(N):= \sum_{k=1}^n {{P(\tau\leq k-1|{\cal F}_{k-1})}\over{P(\tau\leq k|{\cal F}_{k})}}I_{\{\tau<k\}}\Delta N_k+\sum_{k=1}^n I_{\{\tau<k\}}E[\Delta N_k{I}_{\{P(\tau\geq k|{\cal F}_{k})=1\}}|{\cal F}_{k-1}],$$
for any $\mathbb F$-local martingale $N$. This ends the proof of the corollary. 
\end{proof}
\subsection{The case of arbitrary $\mathbb{G}$-martingales}
 This subsection combines Theorem \ref{TheoRepresentation00},  \ref{TheoRepresentation0}, and/or \ref{TheoRepresentation1} with either \cite[Theorems 2.20 and 2.21]{ChoulliDavelooseVanmaele} or \cite[Theorem 2.6]{ChoulliYansori}. These combinations give representation results which extend both \cite{ChoulliDavelooseVanmaele}  and \cite{Azema} to the most general setting. To this end, we recall some results  and notations from  \cite[Theorem 3]{aksamitchoullijeanblanc15} and \cite[Theorems 2.3 and 2.11]{ChoulliDavelooseVanmaele}. 
\begin{theorem}\label{Toperator} The following assertions hold.\\
{\rm{(a)}} For any $M\in{\cal M}_{loc}(\mathbb F)$,  the process
\begin{equation} \label{processMhat}
{\cal T}^{(b)}(M) := M^\tau -{\widetilde{G}}^{-1} I_{\Rbrack 0,\tau\Rbrack} \is [M,m] +  I_{\Rbrack 0,\tau\Rbrack} \is\Big(\sum \Delta M I_{\{\widetilde G=0<G_{-}\}}\Big)^{p,\mathbb F}\end{equation}
 is a $\mathbb G$-local martingale.\\
 {\rm{(b)}}  The process 
\begin{equation} \label{processNG}
N^{\mathbb G}:=D - \widetilde{G}^{-1} I_{\Rbrack 0,\tau\Rbrack} \is D^{o,\mathbb  F}
\end{equation}
is a $\mathbb G$-martingale with integrable variation. Moreover, $H\is N^{\mathbb G}$ is a $\mathbb G$-local martingale with locally integrable variation for any $H$ belonging to
\begin{equation} \label{SpaceLNG}
{\mathcal{I}}^{o,\mathbb{F}}_{loc}(N^{\mathbb G},\mathbb G) := \Big\{K\in \mathcal{O}(\mathbb F)\ \ \big|\quad \vert{K}\vert G{\widetilde G}^{-1} I_{\{\widetilde{G}>0\}}\is D\in{\cal A}_{loc}(\mathbb G)\Big\}.
\end{equation}
\end{theorem}
Furthermore, for any $q\in [1,+\infty)$ and a $\sigma$-algebra ${\cal H}$ on $\Omega\times [0,+\infty)$, we define
\begin{equation}\label{L1(PandD)Local}
L^q\left({\cal H}, P\otimes dD\right):=\left\{ X\ {\cal H}\mbox{-measurable}:\ {E}[\vert X_{\tau}\vert^q I_{\{\tau<+\infty\}}]<+\infty\right\}.\end{equation}
Below, we elaborate our representation results for $\mathbb{G}$-martingales as follows. 
 \begin{theorem}\label{generalGmartingales}
 Suppose that $\tau$ is an honest time, let $M^{\mathbb{G}}$ be a right-continuous-with-left-limits and $\mathbb{G}$-adapted process with $M^{\mathbb{G}}_0=0$, and consider  the $\mathbb{F}$-stopping time $R$ given by
 \begin{equation*}
R:=\inf\left\{t\geq 0:\quad \widetilde{G}_t=0\right\}.\end{equation*}
Then the following assertions hold.\\
 {\rm{(a)}} $M^{\mathbb{G}}$ is a square integrable $\mathbb G$-martingale if and only if there exists a unique quadruplet $$\left(M^{(\mathbb{F},b)}, M^{(\mathbb{F},a)},\varphi^{(o)},\varphi^{(pr)}\right)\in {\cal M}_0^2(\mathbb{F})\times{\cal M}_0^2(\mathbb{F})\times{L}^2\left({\cal{O}}(\mathbb{F}), P\otimes{G}{\widetilde{G}}^{-1}\mbox{dD}\right)\times{L}^2\left(\rm{Prog}(\mathbb{F}), \mbox{P}\otimes\mbox{dD}\right)$$  and satisfying the following properties:
 \begin{align}
 &M^{(\mathbb{F},b)}=(M^{(\mathbb{F},b)})^R,\quad \Delta{M}^{(\mathbb{F},b)}I_{\{\widetilde{G}=0\}}=0,\quad \varphi^{(o)}=\varphi^{(o)}I_{\Lbrack0,R\Lbrack},\quad E[\varphi^{(pr)}_{\tau}\ \big|\ {\cal{F}}_{\tau}]=0\label{Condition4M(b)}\\
& I_{\{G_{-}=1\}}\is M^{(\mathbb{F},a)}\equiv 0,\quad \Delta{M^{(\mathbb{F},a)}}I_{\{\widetilde{G}=1\}}\equiv0.\label{Condition4M(a)}\\
&M^{\mathbb{G}}={{I_{\Rbrack0,\tau\Rbrack}}\over{G_{-}}}\is {\cal{T}}^{(b)}(M^{(\mathbb{F},b)})+\varphi^{(o)}\is N^{\mathbb{G}}+\varphi^{(pr)}\is D+{{I_{\Rbrack\tau,+\infty\Lbrack}}\over{1-G_{-}}}\is{\cal{T}}^{(a)}(M^{(\mathbb{F},a)}).\label{Representation4generalMG}\\
&E\left[\int_0^{\infty}{{I_{\{\widetilde{G}_s>0\}}}\over{{\widetilde{G}}_s}}d[M^{(\mathbb{F},b)},M^{(\mathbb{F},b)}]_s\right]<\infty\quad\mbox{and}\quad E\left[\int_0^{\infty}{{I_{\{\widetilde{G}_s<1\}}}\over{1-{\widetilde{G}}_s}}d[M^{(\mathbb{F},a)},M^{(\mathbb{F},a)}]_s\right]<\infty.\label{Integrability4M(a)M(b)}
 \end{align}
  {\rm{(b)}}  $M^{\mathbb{G}}$ is a $\mathbb G$-local martingale with integrable variation if and only if there exist unique pairs $$\left(M^{(\mathbb{F},b)}, M^{(\mathbb{F},a)}\right)\in ({\cal M}_{0,loc}(\mathbb{F})\cap{\cal{A}}_{loc}(\mathbb{F}))\times ({\cal M}_{0,loc}(\mathbb{F})\cap{\cal{A}}(\mathbb{F})),$$ 
  and $$(\varphi^{(o)},\varphi^{(pr)})\in{I}^{o}_{loc}(N^{\mathbb{G}},\mathbb{G})\times{L}^1\left(\rm{Prog}(\mathbb{F}), P\otimes\mbox{dD}\right)$$  satisfying (\ref{Condition4M(b)}), (\ref{Condition4M(a)}),
  \begin{equation}\label{Integrability4(MF,phi(0))}
E\left[\int_0^{\infty}{{\vert{G}_{-}^{-1}\Delta{M}^{(\mathbb{F},b)}_s+G_s\varphi^{(o)}_s\vert}\over{\widetilde{G}}_s}I_{\{\widetilde{G}_s>0,\ {G}_{-}>0\}}\mbox{dD}^{o,\mathbb{F}}_s+\sum_{s>0}{{\vert{G}_{-}^{-1}\Delta{M}^{(\mathbb{F},b)}_s-\varphi^{(o)}_s\Delta{D}^{o,\mathbb{F}}_s\vert}\over{\widetilde{G}_s}}G_s{I}_{\{\widetilde{G}_s>0,\ {G}_{-}>0\}}\right]<\infty,
\end{equation}
and 
\begin{equation}\label{Representation4generalMG4Variation}
M^{\mathbb{G}}={{I_{\Rbrack0,\tau\Rbrack}}\over{G_{-}^2}}\is {\cal{T}}^{(b)}(M^{(\mathbb{F},b)})+\varphi^{(o)}\is N^{\mathbb{G}}+\varphi^{(pr)}\is D+{{I_{\Rbrack\tau,+\infty\Lbrack}}\over{1-G_{-}}}\is{\cal{T}}^{(a)}(M^{(\mathbb{F},a)}).\end{equation}
 {\rm{(c)}} Suppose that $G>0$ (i.e. $R=+\infty$ $P$-a.s.), and (\ref{MainAssumption4TauBis}) holds. Then $M^{\mathbb{G}}\in {\cal{M}}_{loc}(\mathbb{G})$ if and only if there exists a unique quadruplet $$\left(M^{(\mathbb{F},b)}, M^{(\mathbb{F},a)},\varphi^{(o)},\varphi^{(pr)}\right)\in {\cal M}_{0,loc}(\mathbb{F})\times{\mathbb{M}}_{loc}(\mathbb{F},\tau)\times I^{o}_{loc}(N^{\mathbb{G}},\mathbb{G})\times{L}^1_{loc}\left(\rm{Prog}(\mathbb{F}), P\otimes\mbox{dD}\right)$$  satisfying (\ref{Condition4M(b)}), (\ref{Condition4M(a)}), and 
 \begin{align}
 M^{\mathbb{G}}= {\cal{T}}^{(b)}(M^{(\mathbb{F},b)})+\varphi^{(o)}\is N^{\mathbb{G}}+\varphi^{(pr)}\is D+{{I_{\Rbrack\tau,+\infty\Lbrack}}\over{1-G_{-}}}\is{\cal{T}}^{(a)}(M^{(\mathbb{F},a)}).\label{Representation4generalMG-1}
 \end{align}
 {\rm{(d)}} Suppose that $G>0$ (i.e. $R=+\infty$ $P$-a.s.), and both (\ref{MainAssumption4TauBis}) and (\ref{MainAssumption4Tau}) hold. Then $M^{\mathbb{G}}\in {\cal{M}}_{loc}(\mathbb{G})$ if and only if there exists a unique quadruplet $$\left(M^{(\mathbb{F},b)}, \overline{M}^{(\mathbb{F},a)},\varphi^{(o)},\varphi^{(pr)}\right)\in {\cal M}_{0,loc}(\mathbb{F})\times{\cal M}_{0,loc}(\mathbb{F})\times I^{o}_{loc}(N^{\mathbb{G}},\mathbb{G})\times{L}^1_{loc}\left(\rm{Prog}(\mathbb{F}), P\otimes dD\right)$$  satisfying (\ref{Condition4M(b)}), (\ref{Condition4M(a)}), and 
  \begin{align}
 M^{\mathbb{G}}={\cal{T}}^{(b)}(M^{(\mathbb{F},b)})+\varphi^{(o)}\is N^{\mathbb{G}}+\varphi^{(pr)}\is D+{\cal{T}}^{(a)}( \overline{M}^{(\mathbb{F},a)}).\label{Representation4generalMG-2}
 \end{align}
  \end{theorem}
 \begin{proof} On the one hand, it is clear that, due to \cite[Proposition B.1-(b)]{aksamitetal18}, which claims that $(1-G_{-})^{-1}I_{\{G_{-}<1\}}$ is $\mathbb{F}$-locally bounded when  both (\ref{MainAssumption4TauBis}) and (\ref{MainAssumption4Tau}) hold, we put $$\overline{M}^{(\mathbb{F},a)}:=(1-G_{-})^{-1}I_{\{G_{-}<1\}}\is{M}^{(\mathbb{F},a)},$$
 and conclude that assertion (d) follows immediately from assertion (c). On the other hand, it is clear that assertion (c) is the consequence of applying Theorem \ref{TheoRepresentation0} to $M^{\mathbb{G}}-(M^{\mathbb{G}})^{\tau}$ and \cite[Theorem 2.6]{ChoulliYansori}  to $(M^{\mathbb{G}})^{\tau}$, and combining the obtained results afterwards using 
 \begin{equation}\label{Split4MG}
 M^{\mathbb{G}}=(M^{\mathbb{G}})^{\tau}+M^{\mathbb{G}}-(M^{\mathbb{G}})^{\tau}=:(M^{\mathbb{G}})^{\tau}+\overline{M}^{\mathbb{G}}. \end{equation}
 This proves both assertions (c) and (d). Thus, the rest of this proof focuses on proving assertions (a) and (b) in part 1 and 2 respectively.\\
 {\bf Part 1.} Herein, we prove assertion (a). To this end, we suppose $M^{\mathbb{G}}$ is a square integrable martingale, and hence both $(M^{\mathbb{G}})^{\tau}$ and $\overline{M}^{\mathbb{G}}$ are square integrable martingales. On the one hand, by applying Theorem \ref{TheoRepresentation00}-(a) to $\overline{M}^{\mathbb{G}}$ yields the existence of a unique $M^{(\mathbb{F},a)}\in {\cal M}_{0,loc}(\mathbb{F})$ fulfilling (\ref{Condition4M(a)}), the second condition in (\ref{Integrability4M(a)M(b)}), and 
 \begin{equation}\label{Representation4MbarG}
\overline{M}^{\mathbb{G}}=I_{\Rbrack\tau,+\infty\Lbrack}(1-G_{-})^{-1}\is{\cal{T}}^{(a)}(M^{(\mathbb{F},a)}).
 \end{equation}
On the other hand, by applying \cite[Theorem 2.21]{ChoulliDavelooseVanmaele} to $(M^{\mathbb{G}})^{\tau}$, we get the existence of the unique $\left(M', \varphi^{(o)},\varphi^{(pr)}\right)$ which belongs to ${\cal M}_{0,loc}(\mathbb{F})\times I^{o}_{loc}(N^{\mathbb{G}},\mathbb{G})\times{L}^1_{loc}\left(\mbox{Prog}(\mathbb{F}), P\otimes dD\right)$ satisfying (\ref{Condition4M(b)}) and 
 \begin{equation}\label{Equality4M(b)}
 (M^{\mathbb{G}})^{\tau}={{I_{\Rbrack0,\tau\Rbrack}}\over{G_{-}^2}}\is {\cal{T}}^{(b)}(M')+\varphi^{(o)}\is N^{\mathbb{G}}+\varphi^{(pr)}\is D.
 \end{equation}
By looking closely to the proof of  \cite[Theorem 2.20]{ChoulliDavelooseVanmaele} and using \cite[Theorem 2.17-(c)]{ChoulliDavelooseVanmaele}, we easily conclude that $ (M^{\mathbb{G}})^{\tau}$ is square integrable iff $\varphi^{(pr)})\in{L}^2\left(\mbox{Prog}(\mathbb{F}), P\otimes\mbox{dD}\right)$, and both  $\varphi^{(o)}\is{N}^{\mathbb{G}}$ and $I_{\Rbrack0,\tau\Rbrack}G_{-}^{-2}\is {\cal{T}}^{(b)}(M')$ are square integrable. Direct calculations show that    $\varphi^{(o)}\is{N}^{\mathbb{G}}\in{\cal{M}}^2(\mathbb{G})$ if and only if $ \varphi^{(o)}\in{L}^2\left({\cal{O}}(\mathbb{F}), P\otimes{G}{\widetilde{G}}^{-1}\mbox{dD}\right)$, and $I_{\Rbrack0,\tau\Rbrack}G_{-}^{-2}\is {\cal{T}}^{(b)}(M')\in{\cal{M}}^2(\mathbb{G})$ is equivalent to 
\begin{equation}\label{integrability4M'}
E\left[\int_0^{\infty}{{I_{\{G_{s-}>0\}}I_{\{\widetilde{G}_s>0\}}}\over{G_{s-}^2{\widetilde{G}}_s}}d[M',M']_s \right]=E\left[\int_0^{\tau}{1\over{G_{s-}^4}}d[{\cal{T}}^{(b)}(M'),{\cal{T}}^{(b)}(M')]_s \right]<+\infty.\end{equation}
In virtue of the first and second conditions in (\ref{Condition4M(b)}), which are fulfilled by $M'$, and $\widetilde{G}^{-1}\geq 1$, the above inequality implies that $I_{\{G_{-}>0\}}G_{-}^{-1}$ is $M'$-integrable, and  $M^{(\mathbb{F},b)}:=I_{\{G_{-}>0\}}G_{-}^{-1}\is M'\in{\cal{M}}^2(\mathbb{F})$. Furthermore, by inserting it in  (\ref{integrability4M'}), we deduce that $M^{(\mathbb{F},b)}$ fulfills the first condition in (\ref{Integrability4M(a)M(b)}), as well as the first and second conditions in (\ref{Condition4M(b)}). This proves the existence of the quadruplet $\left(M^{(\mathbb{F},b)}, M^{(\mathbb{F},a)},\varphi^{(o)},\varphi^{(pr)}\right)\in {\cal M}_0^2(\mathbb{F})\times{\cal M}_0^2(\mathbb{F})\times{L}^2\left({\cal{O}}(\mathbb{F}), P\otimes{G}{\widetilde{G}}^{-1}\mbox{dD}\right)\times{L}^2\left(\rm{Prog}(\mathbb{F}), P\otimes dD\right)$ and fulfilling (\ref{Condition4M(b)}), (\ref{Condition4M(a)}), (\ref{Representation4generalMG}) and (\ref{Integrability4M(a)M(b)}). For the reverse sense, it is clear that for any such quadruplet satisfying these conditions, one can easily check that all the four processes $\varphi^{(pr)}\is D$, $\varphi^{(o)}\is N^{\mathbb{G}}$, $I_{\Rbrack0,\tau\Rbrack}G_{-}^{-1}\is {\cal{T}}^{(b)}(M^{(\mathbb{F},b)})$ and $I_{\Rbrack\tau,+\infty\Lbrack}(1-G_{-})^{-1}\is{\cal{T}}^{(a)}(M^{(\mathbb{F},a)}) $ are square integrable $\mathbb{G}$-martingales. This ends the proof of assertion (a).\\
{\bf Part 2.} This part proves assertion (b). Similarly as in part 1, we suppose $ M^{\mathbb{G}}\in {\cal{A}}(\mathbb{G})\cap{\cal{M}}_{loc}(\mathbb{G})$, and notice that this fact implies that both $(M^{\mathbb{G}})^{\tau}$ and $\overline{M}^{\mathbb{G}}$ belong to ${\cal{A}}(\mathbb{G})\cap{\cal{M}}_{loc}(\mathbb{G})$. Then a direct application of Theorem  \ref{TheoRepresentation00}-(b) to $\overline{M}^{\mathbb{G}}$ implies the existence of a unique  $M^{(\mathbb{F},a)}\in {\cal M}_{0,loc}(\mathbb{F})\cap{\cal{A}}(\mathbb{F})$  fulfilling (\ref{Condition4M(a)}), and (\ref{Representation4MbarG}). As the assumption $ M^{\mathbb{G}}\in {\cal{A}}(\mathbb{G})$ yields the uniform integrability of $(M^{\mathbb{G}})^{\tau}$, then we apply \cite[Theorem 2.21]{ChoulliDavelooseVanmaele} to it and deduce the existence of the unique $\left(M', \varphi^{(o)},\varphi^{(pr)}\right)$ which belongs to ${\cal M}_{0,loc}(\mathbb{F})\times{I}^{o}_{loc}(N^{\mathbb{G}},\mathbb{G})\times{L}^1_{loc}\left(\mbox{Prog}(\mathbb{F}), P\otimes dD\right)$ satisfying (\ref{Condition4M(b)}) and (\ref{Equality4M(b)}). By looking closely at the proof of  \cite[Theorem 2.21]{ChoulliDavelooseVanmaele}, one can easily notice that the assumption $ M^{\mathbb{G}}\in {\cal{A}}(\mathbb{G})$ forces $\varphi^{(pr)}$ to belong to ${L}^1\left(\mbox{Prog}(\mathbb{F}), P\otimes dD\right)$, and 
\begin{equation}\label{VariationIntegrable4MF}
\varphi^{(o)}\is{N}^{\mathbb{G}}+I_{\Rbrack0,\tau\Rbrack}G_{-}^{-2}\is {\cal{T}}^{(b)}(M')\in {\cal{A}}(\mathbb{G}).\end{equation}
 As  $\varphi^{(o)}\is{N}^{\mathbb{G}}\in{\cal{A}}_{loc}(\mathbb{G})$, we deduce that  $I_{\Rbrack0,\tau\Rbrack}G_{-}^{-2}\is {\cal{T}}^{(b)}(M')\in{\cal{A}}_{loc}(\mathbb{G})$. This implies that the nondecreasing process $Y:=\sum\vert\Delta{M'}\vert{G}_{-}^{-1}\widetilde{G}^{-1}I_{\{G_{-}>0,\ \widetilde{G}>0\}}$ takes finite values on $\Rbrack0,\tau\Rbrack$, or equivalently $I_{\Rbrack0,\tau\Rbrack}\leq I_{\{ Y<\infty\}}$. Then by taking the $\mathbb{F}$-optional projection, we get $\{\widetilde{G}>0\}\subset\{Y<\infty\}$. Then by combing this with the second condition in (\ref{Condition4M(b)}), we conclude that the process $Y$ has a finite variation, and as a result $M'$ has a finite variation and $I_{\{G_{-}<1\}}G_{-}^{-1}$ is $M'$-integrable. Thus, we put $M^{(\mathbb{F},b)}:=I_{\{G_{-}>0\}}\is M'$ and it is easy to check that $M^{(\mathbb{F},b)}\in {\cal{M}}_{loc}(\mathbb{F})\cap{\cal{A}}_{loc}(\mathbb{F})$  and satisfies the first and second condition of (\ref{Condition4M(b)}). Then by combining \cite[Lemma 2.61-(b), Chapter 2]{Jacod} and  (\ref{VariationIntegrable4MF}), we conclude that the pair $(\varphi^{(o)},M^{(\mathbb{F},b)})$ satisfies (\ref{Integrability4(MF,phi(0))}). This proves the existence of unique pairs $\left(M^{(\mathbb{F},b)}, M^{(\mathbb{F},a)}\right)\in ({\cal M}_{0,loc}(\mathbb{F})\cap{\cal{A}}_{loc}(\mathbb{F}))\times ({\cal M}_{0,loc}(\mathbb{F})\cap{\cal{A}}(\mathbb{F}))$ and $(\varphi^{(o)},\varphi^{(pr)})\in{I}^{o}_{loc}(N^{\mathbb{G}},\mathbb{G})\times{L}^1\left(\rm{Prog}(\mathbb{F}), P\otimes dD\right)$ satisfying (\ref{Condition4M(b)}), (\ref{Condition4M(a)}), (\ref{Integrability4(MF,phi(0))}), and (\ref{Representation4generalMG4Variation}). For the reverse sense, it is easy to verify that, for such quadruplet fulfilling those conditions, the three processes $I_{\Rbrack0,\tau\Rbrack}G_{-}^{-1}\is {\cal{T}}^{(b)}(M^{(\mathbb{F},b)})$, $\varphi^{(pr)}\is D$ and $\varphi^{(o)}\is N^{\mathbb{G}}+I_{\Rbrack\tau,+\infty\Lbrack}(1-G_{-})^{-1}\is{\cal{T}}^{(a)}(M^{(\mathbb{F},a)}) $ belong to  ${\cal{A}}(\mathbb{G})\cap{\cal{M}}_{loc}(\mathbb{G})$. This proves assertion (b), and the proof of the theorem is complete.
 \end{proof}

Besides giving a representation for any  uniformly integrable $\mathbb{G}$-martingale, which extends \cite[Theorems 2.20 or 2.21]{ChoulliDavelooseVanmaele} to any $\mathbb{G}$-martingale uniformly integrable, this theorem also extends \cite[Th\'eor\`eme 3]{Azema} to the most general case, where $\mathbb{F}$ is an arbitrary filtration satisfying the usual conditions, and $\tau$ is an arbitrary honest time which might not be an end of an optional random set, and/or might not avoid $\mathbb{F}$-stopping times. Both assertions (c) and (d) of Theorem \ref{generalGmartingales} can be extended to the case of ${\cal{M}}_{loc}^p(\mathbb{G})$, for any $p\in (1,+\infty)$.
\subsection{Proof of Theorems \ref{TheoRepresentation00}, \ref{TheoRepresentation0} and \ref{TheoRepresentation1}}\label{Subsection4proofs} The proofs of these theorems rely essentially on connecting $\mathbb{G}$-martingales with $\mathbb{F}$-adapted processes having some structures. This fact, which is interesting in itself, is singled out in the following lemma. 

\begin{lemma}\label{MF} Suppose $\tau$ is an honest time and $M^{\mathbb G}\in{\cal{M}}_{loc}(\mathbb{G})$. Then the following assertions hold.\\
{\rm{(a)}} If $M^{\mathbb G}\in{\cal{M}}(\mathbb{G})$ and $\sigma$ is an $\mathbb{F}$-stopping time, then there exists a unique $\mathbb F$-martingale $\overline{M}^{\mathbb F}$ satisfying 
\begin{eqnarray}\label{MFproperties0}
(M^{\mathbb G}-(M^{\mathbb G})^{\tau})^{\sigma}=  \overline{M}^{\mathbb F}{{(I_{\Rbrack\tau,+\infty\Lbrack})^{\sigma}}\over{1-G^{\sigma}}},\quad  \overline{M}^{\mathbb F}=( \overline{M}^{\mathbb F})^{\sigma},\quad\mbox{and}\quad \{\widetilde{G}^{\sigma}=1\}\subset \{\overline{M}^{\mathbb F}= 0\}.
\end{eqnarray}
Here $(I_{\Rbrack\tau,+\infty\Lbrack})^{\sigma}$ is the stopped process (i.e. $(I_{\Rbrack\tau,+\infty\Lbrack})^{\sigma}=I_{\{\sigma>\tau\}}I_{\Rbrack\tau,\sigma\Rbrack}+I_{\{\sigma>\tau\}}I_{\Rbrack\sigma,+\infty\Lbrack}$).\\
{\rm{(b)}}  If (\ref{MainAssumption4TauBis}) holds, then there exists a unique $\mathbb F$-martingale $\overline{M}^{\mathbb F}$ such that the $\mathbb{F}$-predictable process $ \overline{M}^{\mathbb F}_{-}(1-G_{-})^{-1}I_{\{G_{-}<1\}}$ is $\mathbb{F}$-locally bounded, and 
\begin{eqnarray}\label{MFproperties}
M^{\mathbb G}-(M^{\mathbb G})^{\tau}=  \overline{M}^{\mathbb F}{{I_{\Rbrack\tau,+\infty\Lbrack}}\over{1-G}},\quad\mbox{and}\quad \{\widetilde{G}=1\}\subset \{\overline{M}^{\mathbb F}= 0\}.
\end{eqnarray}
{\rm{(c)}} If ${M}\in {\cal{M}}_{loc}(\mathbb{F})$ satisfying $ \{\widetilde G=1\}\subset \{{M}= 0\}$, then 
\begin{eqnarray}\label{mMF}
\left(\sum \Delta{M}\Delta{m}I_{\{\widetilde G=1>G_{-}\}}\right)^{p,\mathbb F}=-{M}_{-}(1-G_{-})^{-1}I_{\{G_{-}<1\}}\is\left(I_{\{\widetilde G=1\}}\is [m,m]\right)^{p,\mathbb F},\end{eqnarray}
 and  ${\cal T}^{(a)}({M})$ --defined via (\ref{T(M)afterTau})-- satisfies 

\begin{eqnarray}\label{Mhat4MF}
{\cal T}^{(a)}({M})= I_{\Rbrack\tau,+\infty\Lbrack}\is{M}+{{I_{\Rbrack\tau,+\infty\Lbrack}}\over{1-\widetilde G}}\is [m,{M}]-{{M_{-}I_{\Rbrack\tau,+\infty\Lbrack}}\over{(1-G_{-})^{2}}}\is\left(I_{\{\widetilde G=1>G_{-}\}}\is [m,m]\right)^{p,\mathbb F}.\end{eqnarray}
\end{lemma}
\begin{proof}  We start this proof by noticing that (\ref{Mhat4MF})  follows immediately from combining (\ref{mMF}) and Theorem \ref{OptionalDecompoTheorem}-(a).  Finally,  due to $ \{\widetilde G=1\}\subset \{M^{\mathbb F}= 0\}$, we easily derive 
 $$\sum \Delta M^{\mathbb F}\Delta{m}I_{\{\widetilde G=1>G_{-}\}}= -\sum M^{\mathbb F}_{-}(1-G_{-})I_{\{\widetilde G=1>G_{-}\}}=-{{M^{\mathbb F}_{-}}\over{1-G_{-}}}I_{\{\widetilde G=1>G_{-}\}}\is [m,m],$$
and hence the proof of assertion (c) is complete. Thus, the remaining part of this proof focuses on proving assertions (a) and (b) in two parts.\\
{\bf Part 1.} Here we prove assertion (a). To this end, we remark that there is no loss of generality in assuming that $(M^{\mathbb{G}})^{\tau}\equiv 0$. 
Thanks to Lemma \ref{GtoFpredictable}-(a) (see also \cite{jeulin80} and \cite{barlow}), there exists an $\mathbb F$-optional process $X$ such that 
$$
M^{\mathbb G}=M^{\mathbb G}I_{\Rbrack\tau,+\infty\Lbrack}=XI_{\Rbrack\tau,+\infty\Lbrack}.$$
It is clear that $X$ is RCLL on ${\Rbrack\tau,+\infty\Lbrack}$, and 
$$
(M^{\mathbb G})^{\sigma}=X^{\sigma}(I_{\Rbrack\tau,+\infty\Lbrack})^{\sigma}.$$
Then consider the process  $\overline{M}^{\mathbb F}:=( ^{o,\mathbb F}((M^{\mathbb G})^{\sigma}))^{\sigma}$, which is a uniformly integrable $\mathbb F$-martingale satisfying 
$$ \overline{M}^{\mathbb F}=X^{\sigma}( ^{o,\mathbb F}((I_{\Rbrack\tau,+\infty\Lbrack})^{\sigma}))^{\sigma}=X^{\sigma} (1-{\widetilde{G}}^{\sigma})\quad\mbox{and}\quad \{\widetilde{G}^{\sigma}=1\}\subset \{\overline{M}^{\mathbb F}=0\}.$$
In the above equality, we used the fact $( ^{o,\mathbb F}((I_{\Rbrack\tau,+\infty\Lbrack})^{\sigma}))^{\sigma}=1-{\widetilde{G}}^{\sigma}$ can be proved easily. Therefore, by combining these remarks with $\widetilde{G}=G$ on $\Rbrack\tau,+\infty\Lbrack$ (see \cite[XX.79]{dellacheriemeyer92} for details), we get 
$$
(M^{\mathbb G})^{\sigma}=X^{\sigma}(I_{\Rbrack\tau,+\infty\Lbrack})^{\sigma}={{\overline{M}^{\mathbb F}}\over{1-\widetilde{G}^{\sigma}}}(I_{\Rbrack\tau,+\infty\Lbrack})^{\sigma}={{\overline{M}^{\mathbb F}}\over{1-G^{\sigma}}}(I_{\Rbrack\tau,+\infty\Lbrack})^{\sigma}.$$
This proves (\ref{MFproperties0}), and the proof of assertion (a) is complete as soon as we prove the uniqueness of $\overline{M}^{\mathbb F}$. To this end, we suppose that there exist two $\mathbb F$-martingales $M$ and $M'$ satisfying (\ref{MFproperties0}), and put $M'':=M-M'$. Hence, we get
$${{M''}\over{1-\widetilde{G}^{\sigma}}} (I_{\Rbrack\tau,+\infty\Lbrack})^{\sigma}=0,\quad\mbox{or equivalently}\quad M''(I_{\Rbrack\tau,+\infty\Lbrack})^{\sigma}=0.$$
Then by taking the $\mathbb F$-optional projection in both sides of the latter equation and stopping at $\sigma$ afterwards, we deduce that $(M-M')(1-\widetilde{G}^{\sigma})=0.$ This implies that $\{\widetilde{G}^{\sigma}<1\}\subset\{M=M'\}$ on the one hand. On the other hand, we have $\{\widetilde{G}^{\sigma}=1\}\subset\{M=M'=0\}$. Thus, we deduce that the two $\mathbb F$-martingales $M$ and $M'$ are indistinguishable. This ends the proof of assertion (a).\\
{\bf Part 2.} Here we prove assertion (b). To this end, we suppose that condition  (\ref{MainAssumption4TauBis}) holds and without loss of generality we assume that $(M^{\mathbb{G}})^{\tau}\equiv 0$. As $M^{\mathbb G}\in {\cal{M}}_{loc}(\mathbb{G})$, there exists a sequence of $\mathbb{G}$-stopping times  $(\sigma_n^{\mathbb{G}})_n$ that increases to infinity and $(M^{\mathbb G})^{\sigma_n^{\mathbb{G}}}$ is a uniformly integrable martingale. On the one hand, in virtue of the assumption (\ref{MainAssumption4TauBis}) and \cite[Proposition B.1-(a)]{aksamitetal18}, we obtain the existence of a sequence of $\mathbb F$-stopping times $(\sigma_n)$ that increases to infinity  almost surely, and 
$$\max(\tau, \sigma_n^{\mathbb{G}})=\max(\sigma_n,\tau),\quad P\mbox{-a.s}.\quad n\geq 1.$$
On the other hand, by applying assertion (a) to each pair $(M^{\mathbb G,n}, \sigma_n)$, where $M^{\mathbb G,n}:=(M^{\mathbb G})^{\sigma_n^{\mathbb{G}}}=(M^{\mathbb G})^{\sigma_n}$, we obtain a sequence of $\mathbb{F}$-martingales $(\overline{M}^{(\mathbb{F},n)})_n$ satisfying
\begin{equation}\label{assertion(a)}
M^{\mathbb G,n}=(M^{\mathbb G})^{\sigma_n^{\mathbb{G}}}={{ \overline{M}^{(\mathbb F,n)}}\over{1-G^{\sigma_n}}}(I_{\Rbrack\tau,+\infty\Lbrack})^{\sigma_n}\quad\quad\mbox{and}\quad \overline{M}^{(\mathbb F,n)}I_{\{\widetilde G^{\sigma_n}=1\}}=0.
\end{equation}
By combining the uniqueness of $\overline{M}^{\mathbb{F},n}$ satisfying (\ref{assertion(a)}) (see assertion (a)) and $M^{\mathbb G,n}= (M^{\mathbb G,k})^{\sigma_n}$ for any $k\geq n$, we deduce that
\begin{equation}\label{Result1}
\overline{M}^{(\mathbb F,n)}= (\overline{M}^{(\mathbb F,k)})^{\sigma_n}\quad\mbox{for any}\quad k\geq n.\end{equation}
 Then we put
 $$\sigma_0:=0,\quad \overline{M}^{\mathbb F, 0}:= 0,\quad\mbox{and}\quad \overline{M}^{\mathbb F}:= \sum_{n=1}^{+\infty}I_{\Rbrack\sigma_{n-1},\sigma_n\Rbrack}\is \overline{M}^{\mathbb F,n}.$$
  As $\sigma_n$ increases to infinity almost surely and by using $( \overline{M}^{\mathbb F})^{\sigma_n}=\sum_{k=1}^nI_{\Rbrack\sigma_{k-1},\sigma_k\Rbrack}\is M^{(\mathbb F,k)}\in{\cal M}_{loc}(\mathbb{F})$  and $({\cal{M}}_{loc}(\mathbb{F}))_{loc}={\cal{M}}_{loc}(\mathbb{F})$, we deduce that $M^{\mathbb F}\in{\cal{M}}_{loc}(\mathbb{F})$. Furthermore,  thanks to (\ref{Result1}), we get 
 $$
 (\overline{M}^{\mathbb F})^{\sigma_n}=\sum_{k=1}^nI_{\Rbrack\sigma_{k-1},\sigma_k\Rbrack}\is \overline{M}^{(\mathbb F,k)}=\sum_{k=1}^n(( \overline{M}^{(\mathbb F,k)})^{\sigma_k}- (\overline{M}^{(\mathbb F,k)})^{\sigma_{k-1}})=\sum_{k=1}^n(\overline{M}^{(\mathbb F,k)}- \overline{M}^{(\mathbb F,k-1)})=\overline{M}^{(\mathbb F,n)}.
 $$
 By combining this with (\ref{assertion(a)}), we derive
\begin{align*}
&\overline{M}^{\mathbb F}I_{\{\widetilde G=1\}} =\lim_{n\longrightarrow+\infty}  (\overline{M}^{\mathbb F})^{\sigma_n}I_{\{\widetilde{G}^{\sigma_n}=1\}} =\lim_{n\longrightarrow+\infty} \overline{M}^{(\mathbb F,n)}I_{\{\widetilde{G}^{\sigma_n}=1\}} \equiv0,\\
& M^{\mathbb G}=\lim_{n\longrightarrow\infty}(M^{\mathbb G})^{\sigma_n^{\mathbb{G}}}=\lim_{n\longrightarrow\infty}{{ \overline{M}^{(\mathbb F,n)}}\over{1-G^{\sigma_n}}}(I_{\Rbrack\tau,+\infty\Lbrack})^{\sigma_n}=\lim_{n\longrightarrow\infty}{{ (\overline{M}^{\mathbb F})^{\sigma_n}}\over{1-G^{\sigma_n}}}(I_{\Rbrack\tau,+\infty\Lbrack})^{\sigma_n}={{\overline{M}^{\mathbb F}}\over{1-G}}I_{\Rbrack\tau,+\infty\Lbrack}.
 \end{align*}
This proves the first part of assertion (b). Now we address the second part, which states that the process $\overline{M}^{\mathbb F}_{-}(1-G_{-})^{-1}I_{\{G_{-}<1\}}$ is $\mathbb{F}$-locally bounded. Thus, as $M^{\mathbb G}$ is RCLL and adapted, we conclude that $M^{\mathbb G}_{-}$ is $\mathbb{G}$-locally bounded, and hence, in virtue of (\ref{MFproperties}), $M^{\mathbb G}_{-}-(M^{\mathbb G}_{-})^{\tau}=\overline{M}^{\mathbb F}_{-}(1-G_{-})^{-1}I_{\Rbrack\tau,+\infty\Lbrack}$ is also $\mathbb{G}$-locally bounded. This yields the existence of a sequence of $\mathbb{G}$-stopping times $\tau^{\mathbb{G}}_n$ that increases to infinity and for all $n\geq 0$, 
\begin{equation}\label{Boundedness1}
{{\overline{M}^{\mathbb F}_{-}}\over{1-G_{-}}}I_{\Rbrack\tau,+\infty\Lbrack}I_{\Rbrack0,\tau^{\mathbb{G}}_n\Rbrack}\leq n.\end{equation}
On the other hand, thanks to the assumption (\ref{MainAssumption4TauBis}) and \cite[Proposition B.1-(a)]{aksamitetal18}, we obtain the existence of a sequence of $\mathbb{F}$-stopping times $\tau^{\mathbb{F}}_n$ that increases to infinity and $\tau^{\mathbb{G}}_n\vee\tau=\tau^{\mathbb{F}}_n\vee\tau$  $P$-a.s.. Thus, by inserting this in (\ref{Boundedness1}), we derive that 
$$
I_{\Rbrack\tau,+\infty\Lbrack}\leq I_{\Sigma},\quad \Sigma:=\left\{{{\overline{M}^{\mathbb F}_{-}}\over{1-G_{-}}}I_{\{G_{-}<1\}}I_{\Rbrack0,\tau^{\mathbb{F}}_n\Rbrack}\leq n\right\}\in{\cal P}(\mathbb{F}).$$
Then by taking the $\mathbb{F}$-predictable projection on both sides of the above inequality, we get $1-G_{-}\leq I_{\Sigma},$
or equivalently 
$$\{G_{-}<1\}\subset \Sigma:=\left\{{{\overline{M}^{\mathbb F}_{-}}\over{1-G_{-}}}I_{\{G_{-}<1\}}I_{\Rbrack0,\tau^{\mathbb{F}}_n\Rbrack}\leq n\right\}.$$
 By combining this with the fact that we always have  $\{G_{-}=1\}\subset \Sigma$, and the fact that $\tau^{\mathbb{F}}_n$ increases to infinity, we deduce that $\overline{M}^{\mathbb F}_{-}(1-G_{-})^{-1}I_{\{G_{-}<1\}}$ is locally bounded. This proves assertion (c),  and the proof of the lemma is complete.\end{proof}
Besides Lemma \ref{MF}, the proof of Theorem \ref{TheoRepresentation0} requires the following two technical lemmas.
\begin{lemma}\label{D0F}Suppose that $\tau$ is an honest time. Then the following assertions hold.\\
{\rm{(a)}} It always holds that 
\begin{eqnarray*}I_{\Rbrack\tau,+\infty\Lbrack}\is D^{o,\mathbb F}\equiv 0.\end{eqnarray*}
{\rm{(b)}} Suppose $M\in {\cal{M}}_{0,loc}(\mathbb{F})$ satisfying $I_{\{G_{-}=1\}}\is M=I_{\{\widetilde{G}=1\}}\Delta{M}={\cal{T}}(M)= 0$. Then $M$ is a null process (i.e $M=0$).\\
{\rm{(c)}} Suppose that (\ref{MainAssumption4TauBis}) and (\ref{MainAssumption4Tau}) hold.  If $M\in {\cal{M}}_{0,loc}(\mathbb{F})$ satisfying
$$MI_{\{\widetilde{G}=1\}}\equiv 0,\quad \mbox{and}\quad I_{\Rbrack\tau,+\infty\Lbrack}\is{M}= -{{M_{-}}\over{1-G_{-}}}I_{\Rbrack\tau,+\infty\Lbrack}\is m,$$
then $M$ is a null process (i.e. $M\equiv 0$).
\end{lemma}
The proof of this lemma is relegated to Appendix \ref{proof4lemmas}, while below we state our second technical lemma. 
A part of the proof of Theorem  \ref{TheoRepresentation0} relies on the following lemma that is interesting in itself.
\begin{lemma}\label{Integrability4M2G}
Let $N\in {\cal{M}}_{loc}(\mathbb{F})$ satisfying (\ref{Representation200}) (i.e. $I_{\{\widetilde{G}=1\}}\Delta{N}=I_{\{G_{-}=1\}}\is{N}=0$), and consider ${\mathbb{M}}_{loc}(\mathbb{F},\tau)$ defined via (\ref{M(F,Tau)}). If we put $H^{\mathbb{G}}:=(1-G_{-})^{-1}I_{\Rbrack\tau,+\infty\Lbrack}$, then the following assertions hold.\\
{\rm{(a)}}   $H^{\mathbb{G}}\in L({\cal{T}}^{(a)}(N),\mathbb{G})$ and $H^{\mathbb{G}}\is{\cal{T}}^{(a)}(N)\in {\cal{M}}^2(\mathbb{G})$ if and only if 
\begin{equation}\label{SqaureIntegrability4N}{E}\left[\int_0^{\infty}{{I_{\{\widetilde{G}_s<1\}}}\over{1-\widetilde{G}_s}}d[N,N]_s\right]<\infty.\end{equation}
{\rm{(b)}}  $N\in{\cal{A}}(\mathbb{F})$ if and only if $H^{\mathbb{G}}$ is ${\cal{T}}^{(a)}(N)$-integrable and $H^{\mathbb{G}}\is{\cal{T}}^{(a)}(N)\in{\cal{A}}(\mathbb{G})$.\\
{\rm{(c)}}  Suppose that (\ref{MainAssumption4TauBis}) holds. Then $N$ belongs to ${\mathbb{M}}_{loc}(\mathbb{F},\tau)$ if and only if $H^{\mathbb{G}}$ is ${\cal{T}}^{(a)}(N)$-integrable and $H^{\mathbb{G}}\is{\cal{T}}^{(a)}(N)\in{\cal{M}}_{loc}(\mathbb{G})$.
\end{lemma}
The proof of the lemma is relegated to Appendix \ref{proof4lemmas}, the rest of this subsection focuses on proving Theorems \ref{TheoRepresentation00}, \ref{TheoRepresentation0} and \ref{TheoRepresentation1}.
\begin{proof}[Proof of Theorem \ref{TheoRepresentation00}] On the one hand, remark that the uniqueness of an $\mathbb{F}$-local martingale $M$, satisfying both (\ref{Representation200}) and (\ref{Representation201}), is a direct consequence of Lemma \ref{D0F}-(b). On the other hand, thanks to Lemma \ref{Integrability4M2G}-(a), if $M\in {\cal{M}}_{loc}(\mathbb{F})$ satisfying (\ref{Representation200}) -(\ref{Representation200bis}) -(\ref{Representation201}), then $(1-G_{-})^{-1}I_{\Rbrack\tau,+\infty\Lbrack}\in L({\cal{T}}^{(a)}(N),\mathbb{G})$ and  $M^{\mathbb{G}}=(1-G_{-})^{-1}I_{\Rbrack\tau,+\infty\Lbrack}\is{\cal{T}}^{(a)}(N)\in {\cal{M}}^2(\mathbb{G})$. This proves one implication of assertion (a), while the reverse implication will be proved in Step 2.1. Similarly, again due to Lemma \ref{Integrability4M2G}-(c), when $M\in {\cal{M}}_{loc}(\mathbb{F})\cap{\cal{A}}(\mathbb{F})$ satisfying (\ref{Representation200}) and (\ref{Representation201}), we deduce that  $(1-G_{-})^{-1}I_{\Rbrack\tau,+\infty\Lbrack}$ is ${\cal{T}}^{(a)}(N)$-integrable and $M^{\mathbb{G}}=(1-G_{-})^{-1}I_{\Rbrack\tau,+\infty\Lbrack}\is{\cal{T}}^{(a)}(N)\in{\cal{A}}(\mathbb{G})$. This proves one sense in the equivalence of assertion (b), while the reserve sense will proved in step 2.2. Thus, the rest of this proof is divided in two steps. \\
 {\bf Step 1.} This step proves the following fact, which holds for any pair $(M^{\mathbb G},N)\in {\cal{M}}_{loc}(\mathbb{G})\times{\cal{M}}_{loc}(\mathbb{F})$.
  \begin{equation}\label{Representation4pair(MG,M)}
 \mbox{If the pair}\  (M^{\mathbb G},N)\ \mbox{satisfies (\ref{MFproperties}), then}\  (1-G_{-})^2\is M^{\mathbb G}=(1-G_{-})\is{\cal{T}}^{(a)}(N)+N_{-}\is{\cal{T}}^{(a)}(m) .\end{equation}
To prove this claim, we suppose that the pair $(M^{\mathbb G},N)$ fulfills (\ref{MFproperties}), which we recall below 
\begin{equation}\label{MG2MF}
M^{\mathbb G}-(M^{\mathbb G})^{\tau}={{N}\over{1-G}} I_{\Rbrack\tau,+\infty\Lbrack}\quad\mbox{and}\quad NI_{\{\widetilde{G}=1\}}=0.\end{equation}
As a result, by combining these properties with $\Lbrack\tau\Rbrack\subset\{\widetilde{G}=1\}$, we deduce that  
\begin{eqnarray}\label{Equality1}
(M^{\mathbb G}-(M^{\mathbb G})^{\tau}) (1-G)=N I_{\Rbrack\tau,+\infty\Lbrack}=N I_{\Lbrack\tau,+\infty\Lbrack}=N \is D+I_{\Rbrack\tau,+\infty\Lbrack}\is{N}=I_{\Rbrack\tau,+\infty\Lbrack}\is{N}.\end{eqnarray}
Thus, we apply the integration by parts formula to the left-hand-side term of (\ref{Equality1}), and use Lemma \ref{D0F}-(a) afterwards to derive  
\begin{align*}
(M^{\mathbb G}-(M^{\mathbb G})^{\tau}) (1-G)&=(1-G_{-}) I_{\Rbrack\tau,+\infty\Lbrack}\is{M}^{\mathbb G}-(M^{\mathbb G}_{-}-M^{\mathbb G}_{\tau}) I_{\Rbrack\tau,+\infty\Lbrack}\is{G}- I_{\Rbrack\tau,+\infty\Lbrack}\is [M^{\mathbb G},G]\\
&=(1-G_{-}) I_{\Rbrack\tau,+\infty\Lbrack}\is{M}^{\mathbb G}-(M^{\mathbb G}_{-}-M^{\mathbb G}_{\tau})  I_{\Rbrack\tau,+\infty\Lbrack}\is{m}- I_{\Rbrack\tau,+\infty\Lbrack}\is[M^{\mathbb G},m].\end{align*}
By combining this latter equality with (\ref{Equality1}), we derive
\begin{align}\label{equaulity20}
&(1-G_{-})I_{\Rbrack\tau,+\infty\Lbrack}\is{N}\nonumber\\
&=(1-G_{-})^2  I_{\Rbrack\tau,+\infty\Lbrack}\is{M}^{\mathbb G}-(M^{\mathbb G}_{-}-M^{\mathbb G}_{\tau})(1-G_{-}) I_{\Rbrack\tau,+\infty\Lbrack}\is{m}- (1-G_{-})I_{\Rbrack\tau,+\infty\Lbrack}\is[M^{\mathbb G},m].
\end{align}
Then this equation allows us to get $[M^{\mathbb G},m]$ as follows
\begin{align*}
&(1-G_{-})I_{\Rbrack\tau,+\infty\Lbrack}\is [N,m]\\
&=\Bigl((1-G_{-})^2-(1-G_{-})\Delta{m}\Bigr)I_{\Rbrack\tau,+\infty\Lbrack}\is [M^{\mathbb G},m]-(M^{\mathbb G}_{-}-M^{\mathbb G}_{\tau})(1-G_{-}) I_{\Rbrack\tau,+\infty\Lbrack}\is[m,m]\\
&=(1-G_{-})(1-\widetilde{G})I_{\Rbrack\tau,+\infty\Lbrack}\is [M^{\mathbb G},m]-(M^{\mathbb G}_{-}-M^{\mathbb G}_{\tau})(1-G_{-}) I_{\Rbrack\tau,+\infty\Lbrack}\is[m,m].
\end{align*}
Thus, by inserting this in (\ref{equaulity20}) and using $(M^{\mathbb G}_{-}-M^{\mathbb G}_{\tau})I_{\Rbrack\tau,+\infty\Lbrack}=N_{-}(1-G_{-})^{-1}I_{\Rbrack\tau,+\infty\Lbrack}$, we obtain
\begin{align}\label{equaulity21}
(1-G_{-})^2I_{\Rbrack\tau,+\infty\Lbrack} dM^{\mathbb G}&=(1-G_{-})I_{\Rbrack\tau,+\infty\Lbrack}d N+{{1-G_{-}}\over{1-\widetilde{G}}} I_{\Rbrack\tau,+\infty\Lbrack}\is [N,m]\nonumber\\
&\quad+N_{-} I_{\Rbrack\tau,+\infty\Lbrack}d{m}+{{N_{-}}\over{1-\widetilde{G}}} I_{\Rbrack\tau,+\infty\Lbrack}\is[m,m] .\end{align}
Put 
\begin{equation}\label{Mmartingale}
M':= (1-G_{-})\is{N}+N_{-}I_{\{G_{-}<1\}}\is m\in {\cal M}_{0,loc}(\mathbb{F}),\end{equation}
and remark that the condition $NI_{\{\widetilde{G}=1\}}=0$ implies that 
\begin{align}\label{Condition4DeltaM'}
\Delta{M'}I_{\{\widetilde G=1\}}&=(1-G_{-})^{-1}I_{\{\widetilde G=1>G_{-}\}}\Delta{N}+N_{-}(1-G_{-})^{-2}I_{\{\widetilde G=1>G_{-}\}}\Delta{m}\nonumber\\
&=-(1-G_{-})^{-1}N_{-}I_{\{\widetilde G=1>G_{-}\}}+N_{-}(1-G_{-})^{-1}I_{\{\widetilde G=1>G_{-}\}}=0.
\end{align}
Hence, it is easy to check that, due to this latter equality,  we have 
$$ I_{\Rbrack\tau,+\infty\Lbrack}\is{M'}+{{ I_{\Rbrack\tau,+\infty\Lbrack}}\over{1-\widetilde{G}}}\is [M',m]={\cal{T}}^{(a)}(M')=(1-G_{-})\is{\cal{T}}^{(a)}(N)+N_{-}\is {\cal{T}}^{(a)}(m).$$
Therefore, by inserting this equality in (\ref{equaulity21}), the claim (\ref{Representation4pair(MG,M)}) follows, and this ends step 1.\\
 {\bf Step 2.} Herein we prove assertions (a) and (b). To this end, remark that if $M^{\mathbb G}\in{\cal{M}}^2(\mathbb{G})$ or $M^{\mathbb G}\in {\cal{A}}(\mathbb{G})$, then $M^{\mathbb G}$ is a uniformly integrable martingale and one can apply Lemma \ref{MF}-(a) to $M^{\mathbb G}$ directly. As result, there exists an $\mathbb{F}$-martingale $N$ such that the pair $(M^{\mathbb G}, N)$ satisfies (\ref{MFproperties}). Hence, in virtue of Step 1 and its proof, we deduce the existence of an $\mathbb{F}$-martingale $N$ such that $M'$ defined in (\ref{Mmartingale}) satisfies 
 \begin{equation}\label{Propertires4M'}
 (1-G_{-})^2 I_{\Rbrack\tau,+\infty\Lbrack}\is M^{\mathbb G} ={\cal{T}}^{(a)}(M'),\quad I_{\{\widetilde{G}=1\}}\Delta{M'}\equiv 0,\quad I_{\{G_{-}=1\}}\is M'\equiv 0.
 \end{equation}
 {\bf Step 2.1.} Here we suppose $I_{\Rbrack\tau,+\infty\Lbrack}\is{M}^{\mathbb G}$ is square integrable, and prove assertion (a). Thus, due to the first and second properties in (\ref{Propertires4M'}),  we conclude that $I_{\Rbrack\tau,+\infty\Lbrack}\is{M}^{\mathbb G}$ is square integrable iff $M'$ satisfies 
 \begin{align}\label{3rdCondition}
{E}\left[\int_0^{\infty}{{I_{\{G_{s-}<1\}}}\over{(1-G_{s-})^2}}\is[M',M']_s\right]&\leq {E}\left[\int_0^{\infty}{{I_{\{G_{s-}<1\}}I_{\{\widetilde{G}_s<1\}}}\over{(1-G_{s-})^2(1-\widetilde{G}_s)}}d[M',M']_s\right]\nonumber\\
&=E \left[\left(I_{\Rbrack\tau,+\infty\Lbrack}\is [M^{\mathbb G} ,M^{\mathbb G} ]\right)_{\infty}\right]<\infty.
 \end{align}
 These (in)equalities are due to the definition of ${\cal{T}}^{(a)}$, the second condition in (\ref{Propertires4M'}) and $1/(1-\widetilde{G})\geq 1$. On the one hand, (\ref{3rdCondition}) implies that $I_{\{G_{s-}<1\}}(1-G_{s-})^{-1}$ is $M'$-integrable (or equivalently $N_{-}I_{\{G_{s-}<1\}}(1-G_{s-})^{-1}$ is $m$-integrable) and 
  $$M:= {{I_{\{G_{s-}<1\}}}\over{1-G_{s-}}}\is M'=I_{\{G_{-}<1\}}\is N+{{N_{-}}\over{1-G_{-}}}I_{\{G_{-}<1\}}\is m\in {\cal{M}}^2(\mathbb{F}).$$
On the other hand, by inserting the above equality in (\ref{3rdCondition}) and (\ref{Propertires4M'}), we obtain
\begin{align*}
&I_{\{G_{-}=1\}}\is M=I_{\{\widetilde{G}_s<1\}}\Delta M\equiv0,\quad  {E}\left[\int_0^{\infty}{{I_{\{\widetilde{G}_s<1\}}}\over{1-\widetilde{G}_s}}d[M,M]_s\right]<\infty, \\
& \mbox{and}\quad (1-G_{-})I_{\Rbrack\tau,+\infty\Lbrack} dM^{\mathbb G}={\cal{T}}^{(a)}(M).\end{align*}
This proves assertion (a), and ends step 2.1.\\
{\bf Step 2.2.} Here we assume that $M^{\mathbb G}\in {\cal{A}}(\mathbb{G})$ and prove assertion (b). Thus, thanks to \cite[Lemma 2.60-(b), Chapter 2]{Jacod} which states that Var$(K)=\sum\vert\Delta{K}\vert$ for any local martingale $K$, and in virtue of (\ref{Propertires4M'}), $I_{\Rbrack\tau,+\infty\Lbrack}\is{M}^{\mathbb G}$ has an integrable variation if and only if $M'\in {\cal{A}}(\mathbb{F})$ and 
\begin{align}\label{VariationCondition}
 {E}\left[\int_0^{\infty}{{I_{\{G_{s-}<1\}}}\over{1-G_{s-}}}d\rm{Var}(M')_s\right]=E \left[\left(I_{\Rbrack\tau,+\infty\Lbrack}\is \rm{Var}(M^{\mathbb G})\right)_{\infty}\right]<\infty.
 \end{align}
This equality is due to $\Delta{M'}I_{\{\widetilde{G}=1\}}=0$. Thus, we deduce that $(1-G_{-})^{-1}I_{\{G_{-}<1\}}$ is $M'$-integrable and $M:=(1-G_{-})^{-1}I_{\{G_{-}<1\}}\is{M'}=I_{\{G_{-}<1\}}\is{M}+N_{-}(1-G_{-})^{-1}I_{\{G_{-}<1\}}\is{m}$  belongs to ${\cal{M}}_{loc}(\mathbb{F})$ satisfying 
$$
I_{\{G_{-}=1\}}\is M=I_{\{\widetilde{G}=1\}}\Delta{M}=0,\quad  (1-G_{-}) I_{\Rbrack\tau,+\infty\Lbrack}\is M^{\mathbb G} ={\cal{T}}^{(a)}(M),\quad\mbox{and}\quad  {E}\left[\rm{Var}(M)_{\infty}\right]<\infty.$$ This proves assertion (b), and the proof of the theorem is complete. \end{proof}
\begin{proof}[Proof of Theorem \ref{TheoRepresentation0}] On the one hand, we remark that uniqueness in assertion (c) follows from  Lemma \ref{D0F}-(c). On the other hand, the implication (b)$\Longrightarrow$(c) follows immediately from putting $$M:=I_{\{G_{-}<1\}}\is{M}^{\mathbb{F}}+M^{\mathbb{F}}_{-}(1-G_{-})^{-1}I_{\{G_{-}<1\}}\is m.$$
Furthermore, (c)$\Longrightarrow$(a) is a direct consequence of Lemma \ref{Integrability4M2G}-(b). Thus, the rest of the proof focuses on proving (a)$\Longrightarrow$(c). To this end, we suppose (\ref{MainAssumption4TauBis}) holds and $I_{\Rbrack\tau,+\infty\Lbrack}\is{M}^{\mathbb{G}}\in{\cal{M}}_{loc}(\mathbb{G})$. Then a direct application of Lemma \ref{MF}-(b), and then step 1 of the proof of Theorem \ref{TheoRepresentation00}  (i.e. the claim (\ref{Representation4pair(MG,M)})) afterwards, we conclude the existence of  an $\mathbb{F}$-martingale ${M}^{\mathbb{F}}$ satisfying
\begin{equation}\label{MF2MGrepresentation}
{M}^{\mathbb{F}}I_{\{\widetilde{G}=1\}}=0,\quad {{{M}^{\mathbb{F}}_{-}}\over{1-G_{-}}}I_{\{G_{-}<1\}}\ \mbox{is $\mathbb{F}$-locally bounded},\end{equation}
and 
\begin{equation*}
(1-G_{-})^2I_{\Rbrack\tau,+\infty\Lbrack}\is M^{\mathbb{G}}=(1-G_{-})\is {\cal{T}}^{(a)}({M}^{\mathbb{F}})+{M}^{\mathbb{F}}_{-}I_{\Rbrack\tau,+\infty\Lbrack}\is {\cal{T}}^{(a)}(m).
\end{equation*}
As $M^{\mathbb{F}}_{-}(1-G_{-})^{-1}I_{\{G_{-}<1\}}$ is locally bonded (see the second property in (\ref{MF2MGrepresentation})), then this process is $m$-integrable and the above equality is equivalent to (\ref{Representation22Bis}). Furthermore, by applying Lemma \ref{Integrability4M2G}-(b) to  $M:=I_{\{G_{-}<1\}}\is {M}^{\mathbb{F}}+{M}^{\mathbb{F}}_{-}(1-G_{-})^{-1}I_{\Rbrack\tau,+\infty\Lbrack}\is{m}$, we deduce that $M\in {\mathbb{M}}_{loc}(\mathbb{F},\tau)$. This proves that assertion (b) holds, and ends the proof of the theorem.
\end{proof}
\begin{proof}[Proof of Theorem \ref{TheoRepresentation1}] On the one hand, we remark that the uniqueness in assertion (b) is a direct consequence of Lemma \ref{D0F}-(c), and the fact that ${\cal{T}}^{(a)}(N)=0$ is equivalent to $I_{\Rbrack\tau,+\infty\Lbrack}\is{N}\equiv 0$ for any $N\in {\cal{M}}_{loc}(\mathbb{F})$ satisfying $I_{\{\widetilde{G}=1\}}\Delta{N}\equiv 0$.  On the other hand,  the proof of (a)$\Longrightarrow$(b) follows immediately from Theorem \ref{TheoRepresentation0}. Thanks to \cite[Proposition B.1-(b)]{aksamitetal18}, under both assumptions (\ref{MainAssumption4TauBis}) and (\ref{MainAssumption4Tau}), the process $(1-G_{-})^{-r}I_{\{G_{-}<1\}}$ is $\mathbb{F}$-locally bounded and $(1-G_{-})^{-r}I_{\Rbrack\tau,+\infty\Lbrack}$ is $\mathbb{G}$-locally bounded for any $r\in (0,\infty)$. Thus, the implication  (c)$\Longrightarrow$(a) is a direct consequence of this latter fact, while (b)$\Longrightarrow$(c) follows from combing the local boundedness of $(1-G_{-})^{-r}I_{\{G_{-}<1\}}$ and putting   
$$M:=(1-G_{-})^{-1}I_{\{G_{-}<1\}}\is \widetilde{M}^{\mathbb{F}}+{{ \widetilde{M}^{\mathbb{F}}_{-}}\over{(1-G_{-})^2}}\is m.$$This ends the proof of the theorem.
\end{proof}
 
 \section{Explicit description of all deflators}\label{Chapter4}
 In this section we parametrize explicitly all deflators for the model  $(S-S^{\tau},\mathbb{G},P)$ in terms of the deflators of a transformed model from $(S,\mathbb F)$. This complements \cite[Theorems 3.2 and 3.4]{ChoulliYansori} and allows us to describe all deflators for the whole model $(S,\mathbb{G},P)$. Thus, we start this section by recalling the mathematical definition of deflators and its local martingale deflator variant. 
 \begin{definition}\label{DeflatorDefinition} Consider the model $(X, \mathbb H, Q)$, where $\mathbb H$ is a filtration, $Q$ is a probability, and $X$ is a $(Q,\mathbb H)$-semimartingale. Let $Z$ be a process.\\
{\rm{(a)}}   We call $Z$ a local martingale deflator for $(X,Q,\mathbb H)$ if $Z_0=1$, $Z>0$ and there exists a real-valued and $\mathbb H$-predictable process $\varphi$ such that $0<\varphi\leq 1$ and both $Z$  and $Z(\varphi\is X)$ are $\mathbb H$-local martingales under $Q$.  Throughout the paper, the set of these local martingale deflators will be denoted by ${\cal Z}_{loc}(X,Q,\mathbb H)$.\\
{\rm{(b)}}   We call $Z$ a deflator for $(X,Q,\mathbb H)$ if $Z_0=1$, $Z>0$ and $Z{\cal E}(\varphi\is X)$ is an $\mathbb H$-supermartingale under $Q$, for any $\varphi\in L(X, \mathbb H)$ such that $\varphi\Delta X\geq -1$. The set of all deflators will be denoted by ${\cal D}(X,Q,\mathbb H)$. When $Q=P$, for the sake of simplicity, we simply omit the probability in notations and terminology.
\end{definition}
The rest of this section is divided into three subsections. The first subsection states our main results on deflators for the model $(S-S^{\tau},\mathbb{G})$ and discusses their importance and the key intermediate results. The second subsection extends the result to the full model $(S,\mathbb{G})$, while the third subsection gives the proof of the principal results in the first subsection.
\subsection{Main results }
 
This section describes explicitly the set of all deflators of $(S-S^{\tau},{\mathbb G})$ in terms of deflators for the initial model $(S, \mathbb F)$. Thus, throughout this section, we assume the following assumptions
\begin{equation}\label{Assumptions4Tau}
\tau\quad\mbox{is a finite honest time such that}\quad G_{\tau}<1\quad P\mbox{-a.s. and}\quad \left\{\widetilde{G}=1>G_{-}\right\}=\emptyset.
\end{equation}

 \begin{theorem}\label{GeneralDefaltorDescription4afterTau}
 Suppose that assumptions (\ref{Assumptions4Tau}) hold, and let  $Z^{\mathbb G}$ be a process such that $(Z^{\mathbb G})^{\tau}\equiv 1$. Then the following assertions are equivalent.\\
{\rm{(a)}} $Z^{\mathbb G}$ is a deflator for $(S-S^{\tau}, \mathbb G)$  (i.e., $Z^{\mathbb G}\in {\cal D}(S-S^{\tau}, \mathbb G)$).\\
{\rm{(b)}} There exists a unique pair $\left(K^{\mathbb F}, V^{\mathbb F}\right)$ such that $K^{\mathbb F}\in {\cal M}_{loc}(\mathbb F)$,  $V^{\mathbb F}$ is an $\mathbb F$-predictable RCLL and nondecreasing process such that $V^{\mathbb F}_0=K^{\mathbb F}_0=0$, ${\cal E}(K^{\mathbb F}){\cal E}(-V^{\mathbb F})\in {\cal D}(I_{\{G_{-}<1\}}\is{S}, \mathbb F)$ and 
\begin{equation}\label{repKG1a}
Z^{\mathbb G}={\cal E}(K^{\mathbb G}){\cal E}(-I_{\Rbrack\tau,+\infty\Lbrack}\is{V}^{\mathbb F})\quad\mbox{where}\quad K^{\mathbb G}= {\cal T}^{(a)}(K^{\mathbb F})+(1-G_{-})^{-1}I_{\Rbrack\tau,+\infty\Lbrack}\is {\cal T}^{(a)}(m).\end{equation}
{\rm{(c)}} There exists a unique $Z^{\mathbb F}\in{\cal D}(I_{\{G_{-}<1\}}\is{S}, \mathbb F)$ such that  \begin{equation}\label{repKGMultiGEneral}
Z^{\mathbb G}={{Z^{\mathbb F}/(Z^{\mathbb F})^{\tau}}\over{{\cal E}(-I_{\Rbrack\tau,+\infty\Lbrack}(1-G_{-})^{-1}\is m)}}.\end{equation}
 \end{theorem}
 The theorem gives two different characterizations for deflators of the model $(S-S^{\tau}, \mathbb G)$. Precisely, assertion (b) characterizes deflators in an additive way, while assertion (c) uses a multiplicative structure. The key idea behind the equivalence between the two characterizations is singled out in the following lemma, which is interesting in itself.
 \begin{lemma}\label{lemmaaftertau} 
Suppose that \eqref{Assumptions4Tau} is fulfilled. Then the following assertions hold.\\
{\rm{(a)}} For any $\mathbb F$-semimartingale $X$, we always have 
\begin{equation}
    \frac{{\cal E}(I_{\Rbrack \tau, \infty \Lbrack} \is X)}{{\cal E} \left(-I_{\Rbrack \tau, \infty \Lbrack} (1-G_{-}\right)^{-1} \is m)} = {\cal E}\Bigl( {\mathcal T}^{(a)}(X) +(1-G_{-})^{-1}I_{\Rbrack \tau, \infty \Lbrack}\is {\mathcal T}^{(a)}(m)\Bigr),
\end{equation}
where $ {\mathcal T}^{(a)}(X)$ is given by 
\begin{equation}\label{Ta(X)}
 {\mathcal T}^{(a)}(X) :=I_{\Rbrack \tau, \infty \Lbrack}\is X+{{I_{\Rbrack \tau, \infty \Lbrack}}\over{1-\widetilde{G}}}\is [X,m].\end{equation}
{\rm{(b)}} For any $K^{\mathbb F} \in {\mathcal M}_{loc} (\mathbb F)$, then
\begin{equation}
    M^{\mathbb G} := \frac {I_{\Rbrack \tau, \infty \Lbrack} \is K^{\mathbb F}}{{\cal E} \left (-I_{\Rbrack \tau, \infty \Lbrack} (1-G_{-})^{-1} \is m  \right)}\in {\cal M}_{loc}(\mathbb{G}).
\end{equation}
{\rm {(c)}} For any $\mathbb {F}$-semimartingales $X$ and $Y$, the following holds.
\begin{equation}
\left[ {\cal T}^{(a)}(X), Y\right] = \left [ X, {\cal T}^{(a)}(Y)\right] = \frac{1-G_{-}}{1-\widetilde G} I_{\Rbrack \tau, \infty \Lbrack} \is \left [ X, Y \right ].
\end{equation}
 \end{lemma}
The proof of this lemma is relegated to Appendix \ref{proof4lemmas}. As a particular case of Theorem  \ref{GeneralDefaltorDescription4afterTau}, we characterize the set of all local martingale deflators for $(S-S^{\tau},\mathbb G)$, denoted by ${\cal Z}_{loc}(S-S^{\tau},\mathbb G)$, as follows.

\begin{theorem}\label{LocalMartingaleDeflator4AfterTau}
 Suppose that assumptions (\ref{Assumptions4Tau}) hold, and let  $K^{\mathbb G}$ be a $\mathbb G$-semimartingale such that $(K^{\mathbb G})^{\tau}\equiv 0$. Then the following assertions are equivalent.\\
{\rm{(a)}} $Z^{\mathbb G}:={\cal E}\left(K^{\mathbb G}\right)$ is a local martingale deflator for $(S-S^{\tau}, \mathbb G)$.\\
{\rm{(b)}} There exists a unique $K^{\mathbb F}\in{\cal M}_{loc}(\mathbb F)$ such that $K^{\mathbb F}_0=0$, ${\cal E}\left(K^{\mathbb F} \right)\in{\cal Z}_{loc}(I_{\{G_{-}<1\}}\is{S},\mathbb F),$ and 
\begin{equation}\label{repKG1}
K^{\mathbb G}={\cal T}^{(a)}({K^{\mathbb F}}) + (1-G_{-})^{-1}I_{\Rbrack\tau,+\infty\Lbrack}\is{\cal T}^{(a)}({m}).\end{equation}
{\rm{(c)}} There exists a unique $ Z^{\mathbb F}\in{\cal Z}_{loc}(I_{\{G_{-}<1\}}\is{S},\mathbb F)$ such that 
\begin{eqnarray}\label{repKGMulti}
Z^{\mathbb G}={{Z^{\mathbb F}/(Z^{\mathbb F})^{\tau}}\over{{\cal E}(-I_{\Rbrack\tau,+\infty\Lbrack}(1-G_{-})^{-1}\is m)}}.
\end{eqnarray}
\end{theorem}
The proof of this theorem will be detailed in Subsection \ref{Proofs4mainresultsofSection4}, while in the rest of this subsection we elaborate the description of the set of all deflators for the full model $(S,\mathbb{G},P)$, by combining Theorems \ref{GeneralDefaltorDescription4afterTau} and \ref{LocalMartingaleDeflator4AfterTau} with Choulli and Yansori \cite[Theorems 3.2 and 3.4]{ChoulliYansori}. 
 \begin{theorem}\label{Deflators4GeneralModel} Suppose that (\ref{Assumptions4Tau}) holds and $G>0$. Consider $N^{\mathbb{G}}$ given by (\ref{processNG}), and let $K^{\mathbb{G}}$ be an arbitrary $\mathbb G$-semimartingale. Then the following assertions hold.\\
 {\rm{(a)}} $Z^{\mathbb{G}}:={\cal{E}}(K^{\mathbb{G}})\in {\cal{D}}(S,\mathbb{G})$ if and only if there exists a quadruplet $(Z^{(\mathbb{F},b)},Z^{(\mathbb{F},a)} ,\varphi^{(o)},\varphi^{(pr)})$ that belongs to $ {\cal{D}}(S,\mathbb{F})\times {\cal{D}}(I_{\{G_{-}<1\}}\is{S},\mathbb{F})\times {\cal I}^{o,\mathbb{F}}_{loc}(N^{\mathbb G},\mathbb G)\times{L}^1_{loc}({\rm{Prog}}(\mathbb F),P\otimes dD)$ such that 
 \begin{equation}\label{Conditions4positivity}
 \varphi^{(pr)}>-1,\quad -{{\widetilde G}\over{ G}}<\varphi^{(o)},\quad \varphi^{(o)}(\widetilde G -G)<\widetilde G,\quad P\otimes dD\mbox{-a.e.,}\quad E\left[\varphi^{(pr)}_{\tau}\ \big|\ {\cal{F}}_{\tau}\right]=0,\ P\mbox{-a.s.}
 \end{equation}
 and 
 \begin{equation}\label{ZG2deflator4F}
 Z^{\mathbb{G}}={{(Z^{(\mathbb{F},b)})^{\tau}}\over{{\cal E}(G_{-}^{-1}\is m)^{\tau}}}{{Z^{(\mathbb{F},a)}/( Z^{(\mathbb{F},a)} )^{\tau}}\over{{\cal E}(-I_{\Rbrack\tau,+\infty\Lbrack}(1-G_{-})^{-1}\is m)}} {\cal{E}}(\varphi^{(o)}\is{N}^{\mathbb{G}}){\cal{E}}(\varphi^{(pr)}\is{D}).
 \end{equation}
 {\rm{(b)}} $Z^{\mathbb{G}}:={\cal{E}}(K^{\mathbb{G}})\in {\cal{Z}}_{loc}(S,\mathbb{G})$ if and only if there exists a quadruplet $(Z^{(\mathbb{F},b)},Z^{(\mathbb{F},a)} ,\varphi^{(o)},\varphi^{(pr)})$ which belongs to $ {\cal{Z}}_{loc}(S,\mathbb{F})\times {\cal{Z}}_{loc}(I_{\{G_{-}<1\}}\is{S},\mathbb{F})\times {\cal I}^{o,\mathbb{F}}_{loc}(N^{\mathbb G},\mathbb G)\times{L}^1_{loc}({\rm{Prog}}(\mathbb F),P\otimes dD)$ and satisfies both conditions (\ref{Conditions4positivity}) and (\ref{ZG2deflator4F}).
 \end{theorem}
 \begin{proof} Remark that ${\cal{E}}(K^{\mathbb{G}})$ is a (local martingale) deflator for $(S,\mathbb{G})$ if and only if  ${\cal{E}}(I_{\Lbrack0,\tau\Rbrack}\is{K}^{\mathbb{G}})$ is a (local martingale) deflator for $(S^{\tau},\mathbb{G})$  and  ${\cal{E}}(I_{\Rbrack\tau,+\infty\Lbrack}\is{K}^{\mathbb{G}})$ is a (local martingale) deflator for $(S-S^{\tau},\mathbb{G})$. Thus, to prove assertion (a) (respectively assertion (b)), we apply Theorem \ref{GeneralDefaltorDescription4afterTau} (respectively Theorem \ref{LocalMartingaleDeflator4AfterTau}) to  ${\cal{E}}(I_{\Rbrack\tau,+\infty\Lbrack}\is{K}^{\mathbb{G}})$ with the model $(S-S^{\tau},\mathbb{G})$ and get $Z^{(\mathbb{F},a)}$ which belongs to ${\cal{D}}(I_{\{G_{-}<1\}}\is{S},\mathbb{F})$ (respectively belongs to ${\cal{Z}}_{loc}(I_{\{G_{-}<1\}}\is{S},\mathbb{F})$) and  
 \begin{equation}\label{Equation4afterTau}
 {{ Z^{\mathbb{G}}}\over{(Z^{\mathbb{G}})^{\tau}}}={\cal{E}}(I_{\Rbrack\tau,+\infty\Lbrack}\is{K}^{\mathbb{G}})={{Z^{(\mathbb{F},a)}/( Z^{(\mathbb{F},a)} )^{\tau}}\over{{\cal E}(-I_{\Rbrack\tau,+\infty\Lbrack}(1-G_{-})^{-1}\is m)}} .
 \end{equation}
Then we apply \cite[Theorem 3.2]{ChoulliYansori}  (respectively \cite[Theorem 3.4)]{ChoulliYansori}) to  ${\cal{E}}(I_{\Lbrack0,\tau\Lbrack}\is{K}^{\mathbb{G}})$ with the model $(S^{\tau},\mathbb{G})$ and obtain the triplet   $(Z^{(\mathbb{F},b)} ,\varphi^{(o)},\varphi^{(pr)})$ which belongs to ${\cal{D}}(S,\mathbb{F})\times {\cal I}^{o,\mathbb{F}}_{loc}(N^{\mathbb G},\mathbb G)\times{L}^1_{loc}({\rm{Prog}}(\mathbb F),P\otimes dD)$ (respectively  ${\cal{Z}}_{loc}(S,\mathbb{F})\times {\cal I}^{o,\mathbb{F}}_{loc}(N^{\mathbb G},\mathbb G)\times{L}^1_{loc}({\rm{Prog}}(\mathbb F),P\otimes dD)$) and satisfies (\ref{Conditions4positivity}) and 
\begin{equation*}
( Z^{\mathbb{G}})^{\tau}={\cal{E}}(I_{\Lbrack0,\tau\Rbrack}\is{K}^{\mathbb{G}})={{(Z^{(\mathbb{F},b)})^{\tau}}\over{{\cal E}(G_{-}^{-1}\is m)^{\tau}}}{\cal{E}}(\varphi^{(o)}\is{N}^{\mathbb{G}}){\cal{E}}(\varphi^{(pr)}\is{D}).
\end{equation*}
Therefore, the proof of (\ref{ZG2deflator4F}) follows immediately from combining  this latter equation with (\ref{Equation4afterTau}), and this ends the proof of theorem.
 \end{proof}
 \subsection{Two particular cases: The jump-diffusion and the discrete-time models}
 This subsection illustrates the main result of  the previous subsection on the cases where $(S,\mathbb{F})$ follows either a jump-diffusion model or a discrete-time model. Thus, we suppose that a standard Brownian motion $W$ and a Poisson process $N$ with intensity $\lambda>0$ are defined on  the probability space $(\Omega, {\cal F}, P)$ and are  independent. Let $\mathbb F$ be the completed and
 right continuous filtration generated by $W$ and $N$. The stock's price process is supposed to have the following dynamics
\begin{equation}\label{SPoisson2}
S_t=S_0 {\cal E} (X)_t,  \  X_t = \sigma\is W_t+  \zeta\is {N}^{\mathbb F}_t + \int_{0}^{t} \mu_s ds, \  {N_t}^{\mathbb F}:=N_t-\lambda t.
\end{equation}
Here $\mu$, $\sigma$ and $\zeta$ are bounded  and $\mathbb F$-predictable processes, and there exists $\delta\in(0,+\infty)$ such that
\begin{equation}\label{parameters2}
 \sigma>0,\quad \zeta>-1,\quad \sigma+\vert\zeta\vert\geq \delta,\ P\otimes dt\mbox{-a.e.}.
 \end{equation}
\begin{theorem}\label{JumpDiffDeflator2} Suppose (\ref{Assumptions4Tau}) holds and $S$ is given by (\ref{SPoisson2})-(\ref{parameters2}). Then the following hold.\\
{\rm{(a)}} $Z^{\mathbb G}$ is a local martingale deflator for $(S-S^{\tau}, \mathbb G)$ with $(Z^{\mathbb G})^{\tau}=1$ if and only if there exists a unique $(\psi_1,\psi_2)\in L^1_{loc}(W,\mathbb F)\times L^1_{loc}(N^{\mathbb F},\mathbb F)$ satisfying 
\begin{equation*}\label{repZG1a}
 Z^{\mathbb G}={{{ \cal E}(\psi_1{I}_{\Rbrack\tau,+\infty\Lbrack}\is W+\psi_2{I}_{\Rbrack\tau,+\infty\Lbrack}\is N^{\mathbb F})}\over{{\cal E}(-(1-G_{-})^{-1}{I}_{\Rbrack\tau,+\infty\Lbrack}\is m)}}\end{equation*}
 and $P\otimes dt\mbox{-a.e.}$ on $(G_{-}<1)$, $(\psi_1,\psi_2)$ satisfies 
 \begin{equation}\label{Condition4(psi1,psi2)}
 \mu+  \psi_1\sigma + \psi_2\zeta\lambda= 0\quad\mbox{and}\quad \psi_2 > -1 .
\end{equation}
{\rm{(b)}} Suppose furthermore that $G>0$. Then $Z^{\mathbb G}$ is a local martingale deflator for $(S, \mathbb G)$ if and only if there exist unique $(\psi^{(1,b)},\psi^{(2,b)})$ and $(\psi^{(1,a)},\psi^{(2,a)})$ which belong to $L^1_{loc}(W,\mathbb F)\times L^1_{loc}(N^{\mathbb F},\mathbb F)$, $(\varphi^{(o)},\varphi^{(pr)})\in {\cal I}^{o,\mathbb{F}}_{loc}(N^{\mathbb G},\mathbb G)\times{L}^1_{loc}({\rm{Prog}}(\mathbb F),P\otimes dD)$ such that\\
(i) $(\psi^{(1,b)},\psi^{(2,b)})$ satisfies  (\ref{Condition4(psi1,psi2)}) $P\otimes dt$-a.e.,\\
(ii) $(\psi^{(1,a)},\psi^{(2,a)})$  satisfies  (\ref{Condition4(psi1,psi2)}) $P\otimes dt$-a.e. on $(G_{-}<1)$,\\
(iii)  $(\varphi^{(o)},\varphi^{(pr)})$ fulfills (\ref{Conditions4positivity}), and 
\begin{equation*}
 Z^{\mathbb G}={{{ \cal E}\left(\psi^{(1,a)}\is W+\psi^{(2,a)}\is N^{\mathbb F}\right)}\over{{\cal E}(-(1-G_{-})^{-1}{I}_{\Rbrack\tau,+\infty\Lbrack}\is m)}}{{{\cal{E}}(\psi^{(1,b)}\is{W}+\psi^{(2,b)}\is{N}^{\mathbb{F}})^{\tau}}\over{{ \cal E}\left(\psi^{(1,a)}\is W+\psi^{(2,a)}\is N^{\mathbb F}\right)^{\tau}}}{{{\cal{E}}(\varphi^{(o)}\is{N}^{\mathbb{G}})}\over{{\cal{E}}(G_{-}^{-1}\is{m})^{\tau}}}{\cal{E}}(\varphi^{(pr)}\is{D}).\end{equation*}

\end{theorem}
The proof follows immediately  from Theorems \ref{LocalMartingaleDeflator4AfterTau} and \ref{Deflators4GeneralModel}  and the fact that for any $M\in {\cal M}_{loc}(\mathbb F)$, there exists a unique pair of $\mathbb F$-predictable processes $(\psi_1,\psi_2)\in  L^1_{loc}(W,\mathbb F)\times L^1_{loc}(N^{\mathbb F},{\mathbb F})$ such that $M=M_0+\psi_1\is W+\psi_2\is N^{\mathbb F}.$

\begin{theorem}\label{DiscreteTimeDeflators}  Suppose that the pair $(\tau,\mathbb{F})$ follows the model given in (\ref{GandG*}), and  
\begin{equation}\label{Condition4TauDiscrete}
P\left(\widetilde{G}_n=1>G_{n-1}\right)=0,\quad n=1,...,T.
\end{equation}
Consider a $\mathbb G$-adapted process $Z^{\mathbb G}$, and the pair  $({\widehat{Q}},\widehat{S})$ given by 
 \begin{equation}\label{Qprobability}
  {\widehat Q}:={\widehat{ Z}}_T\cdot P,\ \mbox{where}\ \ {\widehat Z}_n:=\prod_{k=1}^n \left({{1-\widetilde G_k}\over{1-G_{k-1}}}I_{\{G_{k-1}<1\}}+I_{\{G_{k-1}=1\}}\right),\quad \widehat{S}_n:=\sum_{k=1}^n I_{\{G_{k-1}<1\}}\Delta{S}_k.
 \end{equation}
 Then the following assertions are equivalent.\\
 {\rm{(a)}}  $Z^{\mathbb G}$ is a deflator for $(S-S^{\tau}, \mathbb G)$  (i.e., $Z^{\mathbb G}\in {\cal D}(S-S^{\tau}, \mathbb G)$) with $(Z^{\mathbb G})^{\tau}=1$.\\
 {\rm{(b)}}   There exists a unique $Z\in{\cal D}(\widehat{S}, {\widehat Q},\mathbb F)$ such that $ Z^{\mathbb G}=Z/Z^{\tau}$.
\end{theorem}
\begin{proof} Remark that the process $\widehat{Z}$ is the discrete-time version of the $\mathbb{F}$-local martingale ${\cal{E}}(-(1-G_{-})^{-1}I_{\{G_{-}<1\}}\is m)$. Furthermore, it is easy to check that the process $\widehat{Z}$ is a martingale, and hence the probability $\widehat{Q}$ is well defined. Thus the proof of the theorem follows from combining these remarks with Theorem  \ref{GeneralDefaltorDescription4afterTau}. This ends the proof of the theorem.
\end{proof}
 \subsection{Proof of Theorems \ref{GeneralDefaltorDescription4afterTau} and \ref{LocalMartingaleDeflator4AfterTau}}\label{Proofs4mainresultsofSection4}
 The proof of these theorems requires the following two lemmas that are interesting in themselves. 
 
\begin{lemma}\label{LemmaGTM} The following equalities hold.
 \begin{equation}\label{GTM}
    X^G:=\dfrac{1-G}{1-G^{\tau}}:=I_{ \Lbrack0,\tau\Rbrack}+{{1-G}\over{1-G_{\tau}}} I_{\Rbrack \tau, \infty \Lbrack}= {\cal E}\Bigl(-\frac{1}{1-G_{-}}I_{\Rbrack \tau, \infty \Lbrack} \is m\Bigr)=:{\cal E}({m}^{(a,\mathbb{G})})
 \end{equation}
\end{lemma}
 \begin{proof} Remark that $X_0^G= 1$. Furthermore, by combining $G = m - D^{o, \mathbb F}$ and Lemma \ref{D0F} (i.e. $I_{\Rbrack\tau,+\infty\Lbrack}\is D^{o,\mathbb F} = 0$), we derive  
 \begin{eqnarray*}
     dX_t^G =  I_{\Rbrack \tau ,\infty \Lbrack} (t) dX_t^G + I_{ \Lbrack 0, \tau \Rbrack}(t) dX_t^G  = \dfrac{1}{1-G_{\tau}} I_{\Rbrack \tau, \infty \Lbrack} (t) d(1-G_t)=  \dfrac{-1}{1-G_{\tau}} I_{\Rbrack \tau, \infty \Lbrack}(t) dm_t.
 \end{eqnarray*}
As a result, we conclude that the process $X^G$ satisfies the following SDE
 \begin{eqnarray}
     dX= - \dfrac{X_{-}}{1-G_{-}} I_{\Rbrack \tau, \infty \Lbrack} dm,\quad X_0=1,
 \end{eqnarray}
 which has a unique solution given by the RHS term of \eqref{GTM}. 
 \end{proof}
The second lemma of this subsection connects $\mathbb{G}$-predictable nondecreasing processes with $\mathbb{F}$-predictable nondecreasing processes. 
 \begin{lemma}\label{Lemma1forTheorem4.1} {\rm{(a)}} If $V^{\mathbb{G}}$ is a RCLL, nondecreasing and $\mathbb{G}$-predictable process such that $(V^{\mathbb{G}})^{\tau}\equiv 0$, then there exists a unique RCLL, nondecreasing and $\mathbb{F}$-predictable process $V^{\mathbb{F}}$ such that $V^{\mathbb{F}}_0=0$,
 \begin{equation}\label{VG2VF}
 I_{\{G_{-}=1\}}\is V^{\mathbb{F}}\equiv 0,\quad\mbox{ and}\quad V^{\mathbb{G}}=I_{\Rbrack\tau,+\infty\Lbrack}\is{V}^{\mathbb{F}}.\end{equation}
Furthermore $\Delta{V}^{\mathbb{G}}<1$ if and only $\Delta{V}^{\mathbb{F}}<1$.\\
 {\rm{(b)}} If  (\ref{Assumptions4Tau})  holds, then
 \begin{equation}\label{Equality4SetsTheta}
  \Theta_b(S-S^{\tau},\mathbb{G})=\Bigl\{\varphi{I}_{\Rbrack\tau,+\infty\Lbrack}\quad:\quad \varphi\in   \Theta_b(I_{\{G_{-}<1\}}\is{S},\mathbb{F})\Bigr\}.
\end{equation}
Here, for any filtration $\mathbb{H}$ and any $\mathbb{H}$-semimartingale $X$, the set $\Theta_b(X,\mathbb{H})$ is given by 
   \begin{eqnarray}\label{SpaceL(X,H)}
 {\Theta}_b(X, \mathbb H)  := \left\{ \varphi \mbox{\ is \ } {\mathbb H}\mbox{-predictable and bounded\ }: \quad \varphi \Delta X > -1 \right\}.\end{eqnarray}
 \end{lemma}
 The proof of this lemma is relegated to Appendix \ref{proof4lemmas}. Throughout the rest of the paper,  processes will be compared to each other in the following sense.
 \begin{definition} Let $X$ and $Y$ be two processes such that $X_0=Y_0$. Then we denote 
\begin{eqnarray}\label{IncPro}
 X \succeq Y\quad \mbox{if}\quad X-Y \quad \mbox{is a nondecreasing process.} \end{eqnarray}
 \end{definition}
 Now we are in the stage of delivering the proof of Theorem \ref{GeneralDefaltorDescription4afterTau}.
  \begin{proof}[Proof of Theorem \ref{GeneralDefaltorDescription4afterTau}] The proof is divided into two parts. The first part proves (b) $\Longleftrightarrow$ (c), and the implication (c) $\Longrightarrow$ (a), while the second part focuses on proving (a) $\Longrightarrow$ (b).\\
 {\bf Part 1.} Remark that the implication (b) $\Longrightarrow$ (c) follows directly from Lemma \ref{lemmaaftertau}-(a), while the reverse implication is the consequence of a combination of Lemma \ref{lemmaaftertau}-(a) and the following fact: For any positive $\mathbb{H}$-supermartingale $Z$ with $Z_0=1$, there exists a unique $M\in {\cal M}_{loc}(\mathbb{H})$ and a nondecreasing, RCLL and $\mathbb{H}$-predictable process $V$ such that $V_0=M_0=0$, $\Delta{V}<1$, $\Delta{M}>-1$  and $Z={\cal E}(M){\cal E}(-V)$. For more details about this fact, we refer the reader to \cite[Th\'eor\`eme (6.19)]{Jacod}. This ends the proof of (b) $\Longleftrightarrow$ (c). Thus, the rest of this part proves (c) $\Longrightarrow$ (a). To this end, we assume that assertion (c) holds. Then we notice that, for any $\varphi\in \Theta_b(I_{\{G_{-}<1\}}\is S,\mathbb{F})$, $Z^{\mathbb{F}}{\cal{E}}(\varphi{I}_{\{G_{-}<1\}}\is{S})$ is a positive $\mathbb{F}$-supermartingale, and 
 $$Z^{\mathbb{F}}{\cal{E}}(\varphi{I}_{\{G_{-}<1\}}\is{S})=1+M-V,$$
 where $M\in {\cal M}_{loc}(\mathbb{F})$, $V$ is nondecreasing, RCLL and $\mathbb F$-predictable and $M_0=V_0=0$. Furthermore, direct calculations show that 
 \begin{equation}\label{TrasformedEquation}
 {{Z^{\mathbb{F}}{\cal{E}}(\varphi{I}_{\{G_{-}<1\}}\is{S})}\over{(Z^{\mathbb{F}})^{\tau}{\cal{E}}(\varphi{I}_{\{G_{-}<1\}}\is{S})^{\tau}}}={{Z^{\mathbb{F}}}\over{(Z^{\mathbb{F}})^{\tau}}}{\cal{E}}(\varphi{I}_{\Rbrack\tau,+\infty\Lbrack}\is{S})=1+{I}_{\Rbrack\tau,+\infty\Lbrack}\is{M}-{I}_{\Rbrack\tau,+\infty\Lbrack}\is{V}.\end{equation}
 Then by combining this equality with Lemma \ref{lemmaaftertau}-(b) and the fact that  
 $${{{I}_{\Rbrack\tau,+\infty\Lbrack}\is{V}}\over{{\cal E}({m}^{(a,\mathbb{G})})}}=({I}_{\Rbrack\tau,+\infty\Lbrack}\is{V})\is {1\over{{\cal E}({m}^{(a,\mathbb{G})})}} +{1\over{{\cal E}_{-}({m}^{(a,\mathbb{G})})}}{I}_{\Rbrack\tau,+\infty\Lbrack}\is{V}$$ is a non negative local submartingale, where ${m}^{(a,\mathbb{G})}$ is defined in (\ref{GTM}), we deduce that 
 $${{Z^{\mathbb{F}}/(Z^{\mathbb{F}})^{\tau}}\over{{\cal E}({m}^{(a,\mathbb{G})})}}{\cal{E}}(\varphi{I}_{\Rbrack\tau,+\infty\Lbrack}\is{S})\quad\mbox{is a nonnegative $\mathbb{G}$-local supermartingale }.$$ and hence, this latter process is indeed a $\mathbb{G}$-supermartingale. Thus, the implication (c) $\Longrightarrow$ (a) follows immediately from this latter fact. This ends the first part.\\
 {\bf Part 2.} Here we prove (a) $\Longrightarrow$ (b). To this end, we assume that assertion (a) holds, and hence  $Z^{\mathbb{G}}\in {\cal D}(S-S^{\tau}, \mathbb{G})$ with $(Z^{\mathbb{G}})^{\tau}=1$. Remark that there always exist $M\in {\cal {M}}_{0,loc}(\mathbb{F})$ and a RCLL, $\mathbb{F}$-predictable process with finite variation  $A$ such that  $M_0=A_0=0$,
 \begin{equation}\label{Equation4S}
 S=S_0+M+A+\sum  I_{\{\vert \Delta S\vert>1\}}\Delta S\quad\mbox{and}\quad \max(\vert\Delta{A}\vert,\vert\Delta{M}\vert)\leq 1.\end{equation}
 By applying Proposition \ref{GeneralSupDeflators} to the model $(S-S^{\tau},\mathbb{G})$, we obtain the existence of $M^{\mathbb G}\in {\cal M}_{loc}(\mathbb{G})$ and a RCLL nondecreasing and $\mathbb{G}$-predictable process $V^{\mathbb{G}}$ such that 
 \begin{equation}\label{Condition1}
 Z^{\mathbb{G}}={\cal E}( M^{\mathbb{G}}){\cal E}(-V^{\mathbb{G}}),\quad M_0^{\mathbb{G}}=V_0^{\mathbb{G}}=0,\quad \Delta{V}^{\mathbb{G}}<1\quad\mbox{and}\quad \Delta{M}^{\mathbb{G}}>-1\end{equation}
 and 
 \begin{equation}\label{Condition2}
  \sup_{0<s\leq\cdot}\vert\Delta Y^{(\varphi,\mathbb{G})} \vert\in{\cal A}_{loc}(\mathbb{G}),\quad {1\over{1-\Delta{V}^{\mathbb{G}}}}\is V^{\mathbb{G}} \succeq  A^{(\varphi, M^{\mathbb{G}},\mathbb{G})},\ \forall\ \varphi\in \Theta_b(S-S^{\mathbb{G}}, \mathbb{G}),
 \end{equation}
 where $A^{(\varphi, M^{\mathbb{G}},\mathbb{G})}$ is $\mathbb{G}$-predictable belonging to ${\cal A}_{loc}(\mathbb{G})$ and 
   \begin{equation}\label{Condition2}
   Y^{(\varphi,\mathbb{G})}:= \varphi \is(S-S^{\tau}) + [\varphi \is(S-S^{\tau}), M^{\mathbb{G}}],\quad Y^{(\varphi,\mathbb{G})}-A^{(\varphi, M^{\mathbb{G}},\mathbb{G})}\in{\cal M}_{loc}(\mathbb{G}). \end{equation}
 
 Thus, by applying Theorem \ref{TheoRepresentation1} to $M^{\mathbb{G}}$ and Lemma \ref{Lemma1forTheorem4.1}-(a) to $V^{\mathbb{G}}$, we deduce the existence of a pair $(N^{\mathbb F}, V^{\mathbb{F}})$ such that  $N^{\mathbb F}\in {\cal{M}}_{loc}(\mathbb{F})$ and $V^{\mathbb{F}}$ is RCLL nondecreasing and $\mathbb{F}$-predictable such that 
\begin{align}
  M^{\mathbb G}&={\cal T}^{(a)}(N^{\mathbb F}),\quad I_{\{G_{-}<1\}}\is N^{\mathbb{F}}=0,\quad \Delta N^{\mathbb F}I_{\{\widetilde G=1\}}=0,\quad\Delta{M}^{\mathbb G}={{1-G_{-}}\over{1-\widetilde{G}}}\Delta{N}^{\mathbb F}{I}_{\Rbrack\tau,+\infty\Lbrack}\label{Equation4MG}\\
       V^{\mathbb G}&=I_{\Rbrack\tau,+\infty\Lbrack}\is{V}^{\mathbb F},\quad I_{\{G_{-}=1\}}\is{V}^{\mathbb F}=0\quad\mbox{and}\quad \Delta{V}^{\mathbb F}<1.\label{Equation4VG}
  \end{align}
Thanks to Lemma \ref{Lemma1forTheorem4.1}-(b), there is no loss of generality in considering $\varphi\in \Theta_b(I_{\{G_{-}<1\}}\is{S}, \mathbb{F})$ only. Therefore, for $\varphi\in \Theta_b(I_{\{G_{-}<1\}}\is{S}, \mathbb{F})$, we calculate $[M^{\mathbb G}, \varphi \is (S-S^{\tau})]$ using (\ref{Equation4MG}) and (\ref{Equation4S}) as follows.
\begin{eqnarray*}
&&[  M^{\mathbb G}, \varphi \is (S-S^{\tau})]\\
&=&[  M^{\mathbb G},  \varphi \is(A-A^{\tau})]+  \varphi \is [M^{\mathbb G}, M-M^{\tau}]+\sum\Delta{M}^{\mathbb G}  \varphi \Delta S  I_{\{\vert \Delta S\vert>1\}}I_{\Rbrack\tau,+\infty\Lbrack}\\
&= &\varphi \is[M^{\mathbb G}, A]+\varphi\is [M, {\cal T}^{(a)}(N^{\mathbb F})]+\sum{{1-G_{-}}\over{1-\widetilde{G}}}\Delta N^{\mathbb F}\varphi \Delta S  I_{\{\vert \Delta S\vert>1\}}I_{\Rbrack\tau,+\infty\Lbrack}\\
&=& \varphi \is [{M}^{\mathbb G}, A]+{{1-G_{-}}\over{1-\widetilde{G}}}\varphi{I}_{\Rbrack\tau,+\infty\Lbrack}\is [M, N^{\mathbb F}]+\sum{{1-G_{-}}\over{1-\widetilde{G}}}\Delta N^{\mathbb F}\varphi \Delta S  I_{\{\vert \Delta S\vert>1\}}I_{\Rbrack\tau,+\infty\Lbrack}.\end{eqnarray*}
As both processes $[ M^{\mathbb G}, A]=\Delta A\is M^{\mathbb G}$ and $M-M^{\tau}+(1-G_{-})^{-1}I_{\Rbrack\tau,+\infty\Lbrack}\is \langle M, m\rangle^{\mathbb F}$ are $\mathbb G$-local martingales, due to Yoeurp's lemma (see \cite[Th\'eor\`eme 36, Chapter VII, p. 245]{dellacheriemeyer80}) and Theorem \ref{OptionalDecompoTheorem}-(b) respectively, we derive 
\begin{eqnarray*}
  Y^{(\varphi,\mathbb{G})}&&:=\varphi \is(S-S^{\tau})+[M^{\mathbb G},  \varphi \is(S-S^{\tau})]\\
 &&=-{{\varphi{I}_{\Rbrack\tau,+\infty\Lbrack}}\over{1-G_{-}}} \is \langle M, m\rangle^{\mathbb F} + \varphi \is (A-A^{\tau}) +{{1-G_{-}}\over{1-\widetilde{G}}}\varphi{I}_{\Rbrack\tau,+\infty\Lbrack} \is [M, N^{\mathbb F}]\\
&&+\sum\left(1+{{1-G_{-}}\over{1-\widetilde{G}}}\Delta N^{\mathbb F}\right)\varphi \Delta S  I_{\{\vert \Delta S\vert>1\}}I_{\Rbrack\tau,+\infty\Lbrack}+ {\mathbb G}\mbox{-local martingale}.
\end{eqnarray*}
Therefore, from this equation, we deduce that $ \sup_{0<s\leq\cdot}\vert\Delta Y^{(\varphi,\mathbb{G})} \vert\in{\cal A}_{loc}(\mathbb{G})$ if and only if 
 \begin{equation} W:=\sum\left(1+{{1-G_{-}}\over{1-\widetilde{G}}}\Delta N^{\mathbb F}\right) \varphi\Delta S  I_{\{\vert \Delta S\vert>1\}}I_{\Rbrack\tau,+\infty\Lbrack}\in {\cal A}_{loc}(\mathbb G)\label{integrabilitycondition}\end{equation}
 and in this case we have 
 \begin{equation}
 A^{(\varphi, M^{\mathbb{G}},\mathbb{G})}=-{{\varphi{I}_{\Rbrack\tau,+\infty\Lbrack}}\over{1-G_{-}}} \is \langle M, m\rangle^{\mathbb F} + \varphi{I}_{\Rbrack\tau,+\infty\Lbrack} \is{A} +\varphi{I}_{\Rbrack\tau,+\infty\Lbrack}\is \langle M, N^{\mathbb F}\rangle^{\mathbb F}+W^{p,\mathbb G}.
 \end{equation}
  Remark that, in virtue of Lemma \ref{G2Fcompensator}, $W\in {\cal A}_{loc}(\mathbb G)$  if and only if $U\in {\cal A}_{loc}(\mathbb F)$ and
  \begin{equation}\label{Equation4U}
  W={{I_{\Rbrack\tau,+\infty\Lbrack}}\over{1-\widetilde{G}}}\is{U},\quad\mbox{where}\quad U:=\sum\left(1-\widetilde{G}+(1-G_{-})\Delta N^{\mathbb F}\right) \varphi\Delta{S} I_{\{\vert \Delta S\vert>1\}}I_{\{G_{-}<1\}}.
  \end{equation}
  Put  \begin{eqnarray}\label{NFequation}
K^{\mathbb F}:=N^{\mathbb F}-(1-G_{-})^{-1}I_{\{G_{-}<1\}}\is m\in {\cal M}_{loc}(\mathbb{F}),
\end{eqnarray}  
and on the one hand we obtain 
\begin{equation}\label{KG2KF}
K^{\mathbb G}={\cal{T}}^{(a)}(N^{\mathbb F})={\cal{T}}^{(a)}(K^{\mathbb F})+(1-G_{-})^{-1}I_{\Rbrack\tau,+\infty\Lbrack}\is{\cal{T}}^{(a)}(m).
\end{equation}
This proves (\ref{repKG1a}). On the other hand, we remark that $\Delta{M}^{\mathbb{G}}>-1$ if and only if 
$$\Rbrack\tau,+\infty\Lbrack\subset\left\{{{1-G_{-}}\over{1-\widetilde{G}}}\Delta{N}^{\mathbb{F}}>-1\right\},$$
which is equivalent to 
$$\Rbrack\tau,+\infty\Lbrack\subset\left\{1-{{\Delta{m}}\over{1-G_{-}}}I_{\{G_{-}<1\}}+\Delta{N}^{\mathbb{F}}>0\right\}=\left\{1+\Delta{K}^{\mathbb{F}}>0\right\}.$$
By passing to indicator and taking the $\mathbb{F}$-optional projection, we get 
\begin{equation}\label{GtildeDeltaK(F)}
1-\widetilde{G}\leq I_{\left\{1+\Delta{K}^{\mathbb{F}}>0\right\}},\quad\mbox{which implies that}\quad \{\widetilde{G}<1\}\subset \{1+\Delta{K}^{\mathbb{F}}>0\}.\end{equation}
Due to $\{\widetilde{G}=1>G_{-}\}=\emptyset$, we easily prove that 
$$\{\widetilde{G}=1\}\subset\{\Delta{K}^{\mathbb{F}}=0\}\subset \{1+\Delta{K}^{\mathbb{F}}>0\}.$$
Thus, by combining this latter fact with (\ref{GtildeDeltaK(F)}), we deduce that we always have 
\begin{equation}\label{DeltaK(F)}
\Delta{K}^{\mathbb{F}}>-1.\end{equation}
Direct calculations show that  $U\in {\cal A}_{loc}(\mathbb F)$  if and only if $  \sup_{0<s\leq\cdot}\vert\Delta Y^{(\varphi,\mathbb{F})} \vert\in{\cal A}_{loc}(\mathbb{F})$, where 
$$Y^{(\varphi,\mathbb{F})}:=\varphi{I}_{\{G_{-}<1\}}\is S+[K^{\mathbb{F}},I_{\{G_{-}<1\}}\is S].$$ Furthermore, we derive
\begin{equation}
U=(1-G_{-})\is\sum\left(1+\Delta{K}^{\mathbb{F}}\right) \varphi{I}_{\{G_{-}<1\}}\Delta S  I_{\{\vert \Delta S\vert>1\}}\quad\mbox{and}\quad A^{(\varphi, M^{\mathbb{G}},\mathbb{G})}={I}_{\Rbrack\tau,+\infty\Lbrack}\is A^{(\varphi, K^{\mathbb{F}},\mathbb{F})}.
\end{equation}
By inserting this latter equality and (\ref{Equation4VG}) in the second condition of (\ref{Condition2}), we get
$$
{1\over{1-\Delta{V}^{\mathbb{F}}}}\is V^{\mathbb{F}} \succeq  A^{(\varphi, K^{\mathbb{F}},\mathbb{F})}\quad\mbox{for any}\quad \varphi\in \Theta_b(I_{\{G_{-}<1\}}\is S, \mathbb{F}).$$
Thus, by combining this with (\ref{DeltaK(F)}), we conclude that 
\begin{equation}\label{KFisDefaltor}
Z:={\cal{E}}(K^{\mathbb{F}}){\cal{E}}(-V^{\mathbb{F}})\in {\cal D}(I_{\{G_{-}<1\}}\is S, \mathbb{F}).\end{equation}
Therefore,  assertion (b) follows immediately from combining  (\ref{Equation4VG}), (\ref{KG2KF}) and (\ref{KFisDefaltor}). This ends the second part and the proof of the theorem is complete.\end{proof}
\begin{proof}[Proof of Theorem \ref{LocalMartingaleDeflator4AfterTau}] It is clear that the proof of (b) $\Longleftrightarrow$ (c), and the implication (c) $\Longrightarrow$ (a) follows the same footsteps as in the proof of the corresponding claims in Theorem \ref{GeneralDefaltorDescription4afterTau} (see part 1). Hence, the details for these will be omitted herein and the rest of this proof focuses on (a) $\Longrightarrow$ (b). To this end, we remark that due to Lemma \ref{GtoFpredictable}, for any $\mathbb{G}$-predictable process $\varphi^{\mathbb{G}}$ satisfying $0<\varphi^{\mathbb{G}}\leq 1$, there exists an $\mathbb{F}$-predictable process $\varphi$ such that $0<\varphi\leq 1$ and $\varphi^{\mathbb{G}}I_{\Rbrack\tau,+\infty\Lbrack}=\varphi{I}_{\Rbrack\tau,+\infty\Lbrack}$. Suppose that assertion (a) holds. Hence there exists an $\mathbb {F}$-predictable process $\varphi$ such that $0<\varphi\leq 1$ and ${\cal{E}}(K^{\mathbb{G}})(\varphi{I}_{\Rbrack\tau,+\infty\Lbrack}\is S)$ is a $\mathbb{G}$-local martingale. Then, using It\^o's computation rule, and using the notation and calculations in the proof of Theorem \ref{GeneralDefaltorDescription4afterTau}  part 2, we deduce that $Y^{(\varphi,\mathbb{G})}$ is a $\mathbb{G}$-local martingale, or equivalently $W\in {\cal A}_{loc}(\mathbb{G})$ and 
\begin{eqnarray}\label{NGequation}
0&=& -{{\varphi}\over{1-G_{-}}}I_{\Rbrack\tau,+\infty\Lbrack}\is \langle M, m\rangle^{\mathbb F} +  \varphi \is (A-A^{\tau}) +{{ \varphi}\over{(1-G_{-})^2}} I_{\Rbrack\tau,+\infty\Lbrack}\is \langle M, N^{\mathbb F}\rangle^{\mathbb F}\nonumber\\
&=& \varphi \is (A-A^{\tau}) +{{\varphi}\over{(1-G_{-})^2}}I_{\Rbrack\tau,+\infty\Lbrack}\is \langle M, N^{\mathbb F}-(1-G_{-})\is m\rangle^{\mathbb F}.
\end{eqnarray}  
Thus, we deduce that $U\in {\cal{A}}_{loc}(\mathbb{F})$ and by taking the $\mathbb F$-predictable projections on both sides of the above equality, we get
 \begin{eqnarray}\label{NFequation}
0\equiv \varphi{I}_{\{G_{-}<1\}}\is A + \varphi \is \langle {I}_{\{G_{-}<1\}}\is M, K^{\mathbb F}\rangle^{\mathbb F},\quad K^{\mathbb F}:={{I_{\{G_{-}<1\}}}\over{(1-G_{-})^2}}\is (N^{\mathbb F}-(1-G_{-})\is m).
\end{eqnarray}  
This proves that $Y^{(\varphi,\mathbb{F})}$ is an $\mathbb{F}$-local martingale. Hence, by combining this latter fact with $\Delta{K}^{\mathbb{F}}>  -1$ (which can be proved using similar arguments as in the proof of Theorem \ref{GeneralDefaltorDescription4afterTau} part 2), we deduce that ${\cal{E}}(K^{\mathbb{F}})\in {\cal{Z}}_{loc}(I_{\{G_{-}<1\}}\is{S},\mathbb{F})$, and assertion (b) follows immediately. This ends the proof of the theorem. \end{proof}
 
\appendix
\section{$\mathbb{G}$-processes versus $\mathbb{F}$-processes}
 The lemma below connects $\mathbb{G}$-compensators with $\mathbb{F}$-compensators,  and was given in \cite[Lemma 3.2-(b)]{aksamitetal18}.
 \begin{lemma}\label{G2Fcompensator}
  Suppose that $\tau$ is an honest time. Then for any $\mathbb F$-adapted process $V$ with locally integrable variation, one has 
  \begin{equation}
  I_{\Rbrack \tau, +\infty \Lbrack} \is V^{p, \mathbb G} = 
  I_{\Rbrack \tau, +\infty \Lbrack} (1-G_{-})^{-1} \is \left( (1-\widetilde{G}) \is V \right)^{p, \mathbb F}.
  \end{equation}
 \end{lemma}
 
  We recall the following lemma from \cite[Proposition 5.3]{jeulin80}, see also \cite[Proposition A.5]{aksamitetal18}.
 \begin{lemma}\label{GtoFpredictable}
 Suppose that $\tau$ is an honest time and let $H$ be a process. Then the following hold.\\
 {\rm{(a)}} If $H$ is $\mathbb{G}$-optional, then there exists an $\mathbb{F}$-optional process $H^{\mathbb{F}}$ such that 
 \begin{equation}
     H I_{\Rbrack \tau, \infty \Lbrack}= H^{\mathbb{F}} I_{\Rbrack \tau, \infty \Lbrack}.
 \end{equation}
  {\rm{(b)}} If $H$ is $\mathbb{G}$-predictable, then there exists two $\mathbb{F}$-predictable processes $J$ and $K$ such that 
 \begin{equation}
     H = J I_{\Lbrack 0, \tau \Rbrack} + K I_{\Rbrack \tau, \infty \Lbrack}.
 \end{equation}
 If furthermore $C_1<H\leq C_2$ hold for two constants $C_1<C_2$, then both processes $J$ and $K$ satisfy the same inequalities. 
 \end{lemma}
\section{Characterization of deflators}
Here we recall \cite[Proposition 3.1]{ChoulliYansori}, which is an important result on the characterization of deflators. 
  \begin{proposition}\label{GeneralSupDeflators} 
Let $X$ be an  ${\mathbb H}$-semimartingale and $Z$ be a process. Then the following hold.\\
 {\rm{(a)}} $Z$ is a deflator for $(X, {\mathbb H})$ if and only if there exists a unique pair $(N, V)$ such that $N \in {\cal M}_{loc}(\mathbb H)$, $V$ is nondecreasing RCLL and ${\mathbb H}$-predictable, 
 \begin{eqnarray}
 &&\hskip -1.3cm Z :=Z_0{\cal E}(N){\cal E}(-V), \quad N_0=V_0=0,\quad \Delta N> -1,\quad \Delta V<1,\label{MultiDecompoDeflator}\\
 &&\hskip -1.3cm \sup_{0<s\leq\cdot}\vert\Delta Y^{(\varphi)} \vert\in{\cal A}_{loc}(\mathbb H)\quad\mbox{and}\quad\ {1\over{1-\Delta V}}\is V \succeq  A^{(\varphi, N,\mathbb H)},\ \forall\ \varphi\in \Theta_b(X, \mathbb H).\label{deflatorAssumptions}
  \end{eqnarray}
 Here $Y^{(\varphi)}:= \varphi \is X + [\varphi \is X, N] $, $A^{(\varphi, N,\mathbb H)}\in{\cal A}_{loc}(\mathbb H)$  is $\mathbb H$-predictable such that 
 $Y^{(\varphi)}-A^{(\varphi, N,\mathbb H)}\in{\cal M}_{loc}(\mathbb H)$, and $ {\Theta}_b(X, \mathbb H)$ is defined in (\ref{SpaceL(X,H)}).  \\
 {\rm{(b)}} $Z$ is a local martingale deflator for $(X, {\mathbb H})$ (i.e., $Z\in {\cal Z}_{loc}(X,\mathbb H)$) if and only if there exist a real-valued positive and bounded  $\mathbb H$-predictable process $\varphi$ and a unique $N \in {\cal M}_{loc}(\mathbb H)$ such that $N_0=0$,  \begin{eqnarray}
 && \hskip -1cm Z :=Z_0{\cal E}(N),\quad \Delta N> -1,\quad  \sup_{0<s\leq\cdot}\vert \varphi_s \Delta X_s\vert(1 + \Delta N_s) \in {\cal A}_{loc}(\mathbb H),\label{MartingaleDeflator1}\\
 && \hskip -1cm \varphi \is X + [\varphi \is X, N] \in {\cal M}_{loc}(\mathbb H).\label{MartingaleDeflator2}
  \end{eqnarray}
\end{proposition}
\section{Proofs of Lemmas \ref{D0F}, \ref{Integrability4M2G}, \ref{lemmaaftertau}, and \ref{Lemma1forTheorem4.1}}\label{proof4lemmas}
\begin{proof}[Proof of Lemma \ref{D0F}] This proof is achieved in three parts.\\
{\bf Part 1.} This part proves assertion (a). Thanks to \cite[Proposition 5.1]{jeulin80}, we recall that $\tau$ being an honest time is equivalent to $\widetilde G_{\tau}=1$  $P$-a.s. on $\{\tau<+\infty\}$. Thus, we derive 
\begin{eqnarray*}
E\Bigl[I_{\Rbrack\tau,+\infty\Lbrack}\is D^{o,\mathbb F}_{\infty}\Bigr]&=&E\Bigl[(1-\widetilde G)\is D^{o,\mathbb F}_{\infty}\Bigr]=E\Bigl[(1-\widetilde G_{\tau})I_{\{\tau<+\infty\}}\Bigr]=0.
\end{eqnarray*}
The first equality follows directly from the definition of the optional projection, while the second equality is due to the definition of $D^{o,\mathbb{F}}$, see \cite[D\'efinition 73]{dellacheriemeyer80} with optional process $A= D$. \\
{\bf Part 2.} This part proves assertion (b). Consider $M\in{\cal{M}}_{0,loc}(\mathbb{F})$ such that 
\begin{equation}\label{ConditionNull}
I_{\{G_{-}=1\}}\is{M}=I_{\{\widetilde{G}=1\}}\Delta{M}=0,\end{equation}
 and ${\cal{T}}^{(a)}(M)=0$. Then, in virtue of the second condition in (\ref{ConditionNull}) and the definition of ${\cal{T}}^{(a)}$, we get 
$$0={\cal{T}}^{(a)}(M)= I_{\Rbrack\tau,+\infty\Lbrack}\is{M}+{{I_{\Rbrack\tau,+\infty\Lbrack}}\over{1-\widetilde{G}}}\is[M,m].$$
Thus, this yields to 
$${{1-G_{-}}\over{1-\widetilde{G}}}I_{\Rbrack\tau,+\infty\Lbrack}\is[M,m]=[{\cal{T}}^{(a)}(M),m]=0,$$ and hence $ I_{\Rbrack\tau,+\infty\Lbrack}\is{M}=0$. Therefore, by combining this with the two conditions (\ref{ConditionNull}) and 
$$E\left[(I_{\{\widetilde{G}<1\}}\is [M,M])_{\infty}\right]=E\left[\int_{\tau}^{\infty}{1\over{1-\widetilde{G}}}d[M,M]_s\right]=0,$$
we conclude that $M\equiv 0$. This ends the proof of assertion (b), and  the proof of the lemma is complete. \end{proof}

\begin{proof}[Proof of Lemma \ref{Integrability4M2G}]
Put $H^{\mathbb{G}}:=(1-G_{-})^{-1}I_{\Rbrack\tau,+\infty\Lbrack}$ and consider $N\in {\cal{M}}_{loc}(\mathbb{F})$ satisfying 
$$I_{\{G_{-}=1\}}\is{N}\equiv0\quad\mbox{and}\quad I_{\{\widetilde{G}=1\}}\Delta{N}\equiv 0.$$
Thus, in virtue of the latter condition and the definition of ${\cal{T}}^{(a)}$, we derive 
\begin{equation}\label{equality4HG}
(H^{\mathbb{G}})^2\is [{\cal{T}}^{(a)}(N),{\cal{T}}^{(a)}(N)]={{I_{\Rbrack\tau,+\infty\Lbrack}}\over{(1-G_{-})^2}} \is [{\cal{T}}^{(a)}(N), {\cal{T}}^{(a)}(N)]={{I_{\Rbrack\tau,+\infty\Lbrack}}\over{(1-\widetilde{G})^2}} \is [N, N].\end{equation}
Therefore, due to this equality, we get 
$$E\left[\int_0^{\infty}(H^{\mathbb{G}}_s)^2 d[{\cal{T}}^{(a)}(N),{\cal{T}}^{(a)}(N)]_s\right]=E\left[\int_0^{\infty}{{I_{\{\widetilde{G}_s<1\}}}\over{1-\widetilde{G}_s}}d[N,N]_s\right].$$
Thus, assertion (a) follows immediately from this equality. Assertion (b) is a consequence of a combination of  \cite[Lemma 2.61, Chapter 2]{Jacod} with $\Delta{\cal{T}}^{(a)}(N)=I_{\Rbrack\tau,+\infty\Lbrack}(1-G_{-})\Delta{N}/(1-\widetilde{G})$, and
$$E\left[\int_0^{\infty}H^{\mathbb{G}}_s d\rm{Var}_s({\cal{T}}^{(a)}(N))\right]=E\left[\sum_{s>0}{H}^{\mathbb{G}}_s\vert\Delta{\cal{T}}^{(a)}_s(N)\vert\right]=E\left[\sum_{s> 0}{{\vert\Delta{N}_s\vert}\over{1-\widetilde{G}_s}}I_{\{s>\tau\}}\right]=E\left[\rm{Var}_{\infty}(N)\right].$$
The rest of this proof focuses on proving assertion (c).  It is well known that $H^{\mathbb{G}}$ is ${\cal{T}}^{(a)}(N)$-integrable and $H^{\mathbb{G}}\is{\cal{T}}^{(a)}(N)\in {\cal{M}}_{loc}(\mathbb{G})$ if and only if $\sqrt{(H^{\mathbb{G}})^2\is [{\cal{T}}^{(a)}(N),{\cal{T}}^{(a)}(N)]}\in {\cal{A}}^+_{loc}(\mathbb{G})$, which is equivalent to 
\begin{equation}\label{Integrability4HG1}
\sqrt{I_{\Rbrack\tau,+\infty\Lbrack}(1-\widetilde{G})^{-2} \is [N, N]}\in{\cal{A}}^+_{loc}(\mathbb{G}),\end{equation}
due to (\ref{equality4HG}). Then by combining $[N,N]=[N^c,N^c]+\sum(\Delta{N})^2$, where $N^c$ is the continuous local martingale part of $N$, and \cite[Proposition 2.56 and Corollary 2.57-(a), Chapter 2]{Jacod}, we deduce that (\ref{Integrability4HG1}) is equivalent to 
$$
{{I_{\Rbrack\tau,+\infty\Lbrack}}\over{(1-G_{-})^2}} \is [N^c, N^c]+ \sum {{(1-\widetilde{G})^{-2}(\Delta{N})^2}\over{1+(1-\widetilde{G})^{-1}\vert\Delta{N}\vert}}I_{\Rbrack\tau,+\infty\Lbrack}\in{\cal{A}}^+_{loc}(\mathbb{G}).$$
Thanks to \cite[Lemma 3.2-(c)]{aksamitetal18}, this latter condition is equivalent to 
$$
{{I_{\{G_{-}<1\}}}\over{1-G_{-}}} \is [N^c, N^c]+ \sum {{(\Delta{N})^2I_{\{\widetilde{G}<1\}}}\over{1-\widetilde{G}+\vert\Delta{N}\vert}}={{I_{\{\widetilde{G}<1\}}}\over{1-\widetilde{G}+\vert\Delta{N}\vert}}\is [N,N]\in{\cal{A}}^+_{loc}(\mathbb{F}),$$
or equivalently $N\in {\mathbb{M}}_{loc}(\mathbb{F},\tau)$. This proves assertion (c), and ends the proof of the lemma.\end{proof}
\begin{proof} [Proof of Lemma \ref{lemmaaftertau}]
1) Here we prove assertion (c). Thanks to (\ref{Assumptions4Tau}) and $\Rbrack \tau, \infty \Lbrack\subset\{G= \widetilde{G}\}$ (for details about this latter fact, we refer the reader to \cite[XX.79]{dellacheriemeyer92}), we derive
\begin{equation}\label{aftertauwithcondition}
    {\cal T}^{(a)}(X)= I_{\Rbrack\tau,+\infty\Lbrack}\is X
 +{1 \over{1- \widetilde{G}}} {I_{\Rbrack\tau,+\infty\Lbrack}} \is [m,X].
 \end{equation} 
Hence, direct calculation yields
\begin{align*}
 \left[ {\cal T}^{(a)}(X), Y\right] & = \left [ I_{\Rbrack\tau,+\infty\Lbrack} \is X + \frac{1}{1-\widetilde{G}} I_{\Rbrack \tau, \infty \Lbrack} \is [m, X], Y \right ]=  I_{\Rbrack\tau,+\infty\Lbrack}\is [X, Y] + \frac{1}{1-\widetilde{G}} I_{\Rbrack \tau, \infty \Lbrack} \Delta m \is [X, Y]\\
 & = \left ( 1 + \frac{\widetilde{G}- G_{-}}{1-\widetilde{G}} \right ) I_{\Rbrack\tau,+\infty\Lbrack} \is [X, Y] = \frac{1-G_{-}}{1-\widetilde G} I_{\Rbrack \tau, \infty \Lbrack} \is \left [ X, Y \right ] = \left [ X, {\cal T}^{(a)}(Y)\right] .   
\end{align*}
This proves assertion (c).\\
2) To prove assertion (a) , we recall that $$1/{\cal E}(X) = {\cal E}\left (-X + (1+ \Delta X)^{-1} \is [X, X]\right ),$$ holds for any semimartingale $X$ such that $1+ \Delta X > 0$, and this fact is a consequence of Yor's formula. Then, by combining this equality and $\Delta m = \widetilde{G}- G_{-}$, we derive
\begin{align}
  \frac{1}{{\cal E} \left(-I_{\Rbrack \tau, \infty \Lbrack} (1-G_{-}\right)^{-1} \is m)} & = {\cal E} \left( \frac{1}{1-G_{-}} I_{\Rbrack \tau , \infty \Lbrack} \is m + \frac{(1-G_{-})^{-2}}{1- \frac{\id_{\Rbrack \tau, \infty \Lbrack} \Delta m}{1-G_{-}}} I_{\Rbrack \tau, \infty \Lbrack} \is [m, m] \right)\nonumber \\
  & = {\cal E} \left( \frac{I_{\Rbrack \tau, \infty \Lbrack} }{1-G_{-}}\is m + \frac{I_{\Rbrack \tau, \infty \Lbrack} }{(1-G_{-})(1-\widetilde{G})}  \is [m, m] \right)\nonumber\\
  & = {\cal E}\left ((1-G_{-})^{-1} I_{\Rbrack \tau, \infty \Lbrack}  \is {\mathcal T}^{(a)}(m)\right ) .\label{1/E(m)}
\end{align}
Therefore, by using this equality and Yor's formula afterwards, for any $X$ we obtain 
\begin{eqnarray*}
 &&\frac { {\cal E} \left(I_{\Rbrack \tau, \infty \Lbrack} \is X \right)}{{\cal E} \left (-I_{\Rbrack \tau, \infty \Lbrack} \frac{1}{1-G_{-}} \is m  \right)} = {\cal E} \left(I_{\Rbrack \tau, \infty \Lbrack} \is X\right) {\cal E}\left ((1-G_{-})^{-1} I_{\Rbrack \tau, \infty \Lbrack}  \is {\mathcal T}^{(a)}(m)\right )\\
&& = {\cal E} \left(I_{\Rbrack \tau, \infty \Lbrack} \is X + (1-G_{-})^{-1} I_{\Rbrack \tau, \infty \Lbrack}\is {\mathcal T}^{(a)}(m) +  \frac{I_{\Rbrack \tau, \infty \Lbrack} }{1-G_{-}} \is \left [ X, {\cal T}^{(a)}(m)\right]  \right)  \\
&& = {\cal E} \left(I_{\Rbrack \tau, \infty \Lbrack} \is X + (1-G_{-})^{-1} I_{\Rbrack \tau, \infty \Lbrack} \is {\mathcal T}^{(a)}(m) +  \frac{I_{\Rbrack \tau, \infty \Lbrack} }{1-\widetilde{G}} \is \left [ X,  m \right]  \right) \\
&& = {\cal E} \left ({\cal T}^{(a)}(X)+(1-G_{-})^{-1} I_{\Rbrack \tau, \infty \Lbrack} \is {\cal T}^{(a)}(m) \right ).\end{eqnarray*}
The third equality above follows from assertion (c). This ends the proof of assertion (a).\\
3) Here we prove assertion (b). Due to the integration by part and (\ref{1/E(m)}), we get 
\begin{eqnarray*}
    && M^{\mathbb G}:= \frac {I_{\Rbrack \tau, \infty \Lbrack} \is K^{\mathbb F}}{{\cal E} \left (-I_{\Rbrack \tau, \infty \Lbrack} (1-G_{-})^{-1} \is m  \right)} \\
    &&= \frac{(I_{\Rbrack \tau, \infty \Lbrack} \is K)_{-}^{\mathbb F}}{(1-G_{-}){\cal E}_{-}(-I_{\Rbrack \tau, \infty \Lbrack} (1-G_{-})^{-1} \is m)} \is {\cal T}^{(a)}(m) +\frac{I_{\Rbrack \tau, \infty \Lbrack}}{{\cal E}_{-}(-I_{\Rbrack \tau, \infty \Lbrack} (1-G_{-})^{-1} \is m)} \is K^{\mathbb F} \\
    && + \frac{I_{\Rbrack \tau, \infty \Lbrack}}{(1-G_{-}){\cal E}_{-}(-I_{\Rbrack \tau, \infty \Lbrack} (1-G_{-})^{-1} \is m)} \is \left [K^{\mathbb F}, {\cal T}^{(a)}(m) \right]\\
    && = \frac{(I_{\Rbrack \tau, \infty \Lbrack} \is K)_{-}^{\mathbb F}}{(1-G_{-}){\cal E}_{-}(-I_{\Rbrack \tau, \infty \Lbrack} (1-G_{-})^{-1} \is m)} \is {\cal T}^{(a)}(m) +\frac{I_{\Rbrack \tau, \infty \Lbrack}}{{\cal E}_{-}(-I_{\Rbrack \tau, \infty \Lbrack} (1-G_{-})^{-1} \is m)} \is K^{\mathbb F} \\
    && + \frac{I_{\Rbrack \tau, \infty \Lbrack}}{(1-\widetilde{G}){\cal E}_{-}(-I_{\Rbrack \tau, \infty \Lbrack} (1-G_{-})^{-1} \is m)} \is \left [K^{\mathbb F}, m \right]\\
    &&= \frac{(I_{\Rbrack \tau, \infty \Lbrack} \is K)_{-}^{\mathbb F}}{(1-G_{-}){\cal E}_{-}(-I_{\Rbrack \tau, \infty \Lbrack} (1-G_{-})^{-1} \is m)} \is {\cal T}^{(a)}(m) +\frac{I_{\Rbrack \tau, \infty \Lbrack}}{{\cal E}_{-}(-I_{\Rbrack \tau, \infty \Lbrack} (1-G_{-})^{-1} \is m)} \is {\cal T}^{(a)}(K^{\mathbb F}) \\
\end{eqnarray*}
Thus, in virtue of Theorem \ref{OptionalDecompoTheorem}, this proves $M^{\mathbb G} \in {\mathcal M}_{loc}(\mathbb G)$, and the proof of the lemma is complete.
\end{proof}
 \begin{proof}[Proof of Lemma \ref{Lemma1forTheorem4.1}] This proof has two parts where we prove assertions (a) and (b) respectively.\\
 {\bf Part 1.} We start proving the uniqueness of the process $V^{\mathbb{F}}$ satisfying (\ref{VG2VF}). This follows from the fact that if $V$ is an $\mathbb{F}$-predictable process with finite variation such that
 \begin{equation}\label{Uniqueness4V}
 V_0=0,\quad I_{\{G_{-}=1\}}\is V=0\quad \mbox{and}\quad I_{\Rbrack\tau,+\infty\Lbrack}\is{V}=0,\end{equation}
 then $V\equiv 0$. To prove this latter fact, we take the dual predictable projection in both sides of the third condition and get $(1-G_{-})\is V=0$, or equivalently $I_{\{G_{-}<1\}}\is V\equiv 0$. Thus, by combine this with the second condition in (\ref{Uniqueness4V}), we deduce that $V=I_{\{G_{-}=1\}}\is V+I_{\{G_{-}<1\}}\is V=0$. This proves the uniqueness of $V^{\mathbb{F}}$. To prove the last statement of assertion (a),  we remark that due to the second equality in (\ref{VG2VF}) we get $\Delta{V}^{\mathbb{G}}=I_{\Rbrack\tau,+\infty\Lbrack}\Delta{V}^{\mathbb{F}}$. Thus, $\Delta{V}^{\mathbb{G}}<1$  if and only if $ I_{\Rbrack\tau,+\infty\Lbrack}\leq I_{\{\Delta{V}^{\mathbb{F}}<1\}}$. By taking the $\mathbb{F}$-predictable projection on both sides of the latter inequality, we get 
 $$1-G_{-}\leq{I}_{\{\Delta{V}^{\mathbb{F}}<1\}},$$
 or equivalently $\{ G_{-}<1\}\subset\{\Delta{V}^{\mathbb{F}}<1\}$. By combining this with 
 $$\{G_{-}=1\}\subset\{\Delta{V}^{\mathbb{F}}=0\}\subset \{\Delta{V}^{\mathbb{F}}<1\},$$ we conclude that $\Delta{V}^{\mathbb{F}}<1$ always hold. This proves that $\Delta{V}^{\mathbb{G}}<1$ implies $\Delta{V}^{\mathbb{F}}<1$, while the reverse inclusion is obvious from the second equality in (\ref{VG2VF}). This ends the proof of the last statement of assertion (a). Thus, the rest of  this part focuses on the existence of the process $V^{\mathbb{F}}$ satisfying $V^{\mathbb{F}}_0=0$ and (\ref{VG2VF}). To this end, remark that there is no loss of generality in assuming that the process $V^{\mathbb{G}}$ is bounded. Thus, in virtue of \cite[Th\'eor\`eme 47, p.119 and Th\'eor\`eme 59, 268]{dellacheriemeyer80} and the nondecreasiness of $V^{\mathbb{G}}$, the process $S^{\mathbb{F}}:=^{o,\mathbb{F}}(V^{\mathbb{G}})$ is a RCLL  and bounded $\mathbb{F}$-submartingale. Thus, on the one hand, we deduce the existence of $M\in {\cal{M}}_{loc}(\mathbb{F})$ and a RCLL nondecreasing and $\mathbb{F}$-predictable process $U$ such that  
 \begin{equation}\label{Projection4Vg}
 S^{\mathbb{F}}:={^{o,\mathbb{F}}(V^{\mathbb{G}})}=S^{\mathbb{F}}_0+M+U,\quad\mbox{and}\quad M_0=U_0=0.
 \end{equation}
 On the other hand, as $V^{\mathbb{G}}$ is a $\mathbb{G}$-predictable process such that $(V^{\mathbb{G}})^{\tau}\equiv 0$, we apply Lemma \ref{GtoFpredictable} and get the existence of an $\mathbb{F}$-predictable process $V$  such that 
 \begin{equation}\label{Vg2V}
 V^{\mathbb{G}}=V^{\mathbb{G}}I_{\Rbrack\tau,+\infty\Lbrack}=VI_{\Rbrack\tau,+\infty\Lbrack}.
 \end{equation}
 By taking the $\mathbb{F}$-optional projection on both sides of this equality, we obtain $S^{\mathbb{F}}=V(1-\widetilde{G})$, which yields $\{\widetilde{G}=1\}\subset\{S^{\mathbb{F}}=0\}$. Thus, by combing this fact with (\ref{Vg2V}), (\ref{Projection4Vg}) and Lemma \ref{LemmaGTM}, we derive 
 \begin{eqnarray}
V^{\mathbb{G}}&&=VI_{\Rbrack\tau,+\infty\Lbrack}={{S^{\mathbb{F}}}\over{1-\widetilde{G}}}I_{\Rbrack\tau,+\infty\Lbrack}={{S^{\mathbb{F}}}\over{1-G}}I_{\Rbrack\tau,+\infty\Lbrack}={{S^{\mathbb{F}}I_{\Rbrack\tau,+\infty\Lbrack}}\over{(1-G_{\tau})X^G}}\nonumber\\
  &&={{I_{\Rbrack\tau,+\infty\Lbrack}\is{S}^{\mathbb{F}}}\over{(1-G_{\tau})X^G}}={{I_{\Rbrack\tau,+\infty\Lbrack}\is{M}}\over{(1-G_{\tau})X^G}}+{{I_{\Rbrack\tau,+\infty\Lbrack}\is{U}}\over{(1-G_{\tau})X^G}}\nonumber\\
  &&={{(I_{\Rbrack\tau,+\infty\Lbrack}\is{M})(1-G_{\tau})^{-1}}\over{X^G}}+(I_{\Rbrack\tau,+\infty\Lbrack}\is{U})\is {{(1-G_{\tau})^{-1}}\over{X^G}}+{{(1-G_{\tau})^{-1}}\over{X^G_{-}}}I_{\Rbrack\tau,+\infty\Lbrack}\is{U}\nonumber\\
  &&=\underbrace{{{(I_{\Rbrack\tau,+\infty\Lbrack}\is{M})(1-G_{\tau})^{-1}}\over{X^G}}+(I_{\Rbrack\tau,+\infty\Lbrack}\is{U})\is {{(1-G_{\tau})^{-1}}\over{X^G}}}_{\mbox{is a $\mathbb{G}$-local martingale}}+
  {1\over{1-G_{-}}}I_{\Rbrack\tau,+\infty\Lbrack}\is{U}.\label{VG2V}
 \end{eqnarray}
Therefore, as  both processes $V^{\mathbb{G}}$  and $ (1-G_{-})^{-1}I_{\Rbrack\tau,+\infty\Lbrack}\is{U}$ are nondecreasing  and $\mathbb{G}$-predictable, we conclude that the $\mathbb{G}$-local martingale part in (\ref{VG2V}) is null. Hence, we obtain 
$$V^{\mathbb{G}}=  {{I_{\Rbrack\tau,+\infty\Lbrack}}\over{1-G_{-}}}\is{U}.$$
Therefore,  by putting $V^{\mathbb{F}}=(1-G_{-})^{-1}I_{\{G_{-}<1\}}\is U$, the proof of assertion (a) is complete. \\
 {\bf Part 2.} Due to Lemma \ref{GtoFpredictable}, remark that $\varphi^{\mathbb{G}}\in \Theta_b(S-S^{\tau},\mathbb{G})$ if and only if there exists an $\mathbb{F}$-predictable and bounded process $\varphi$ such that 
\begin{equation*}
\varphi{I}_{\Rbrack\tau,+\infty\Lbrack}=\varphi^{\mathbb{G}}I_{\Rbrack\tau,+\infty\Lbrack}.
\end{equation*}
 Thus, due to $\Rbrack\tau,+\infty\Lbrack\subset\{G_{-}<1\}$, the above equality implies that 
 $$
 \varphi{I}_{\{G_{-}<1\}}\Delta{S}>-1\quad\mbox{on}\quad\Rbrack\tau,+\infty\Lbrack,\quad\mbox{or equivalently}\quad I_{\Rbrack\tau,+\infty\Lbrack}\leq I_{\{ \varphi{I}_{\{G_{-}<1\}}\Delta{S}>-1\}}. $$
 Then by taking the $\mathbb{F}$-optional projection in both sides of the latter inequality, we get 
 $$1-\widetilde{G}\leq I_{\{ \varphi{I}_{\{G_{-}<1\}}\Delta{S}>-1\}},\quad \mbox{or equivalently}\quad\{\widetilde{G}<1\}\subset \{ \varphi{I}_{\{G_{-}<1\}}\Delta{S}>-1\}.$$
 By combing this latter inclusion with the fact that $\{\widetilde{G}=1\}\subset\{G_{-}<1\}\subset \{ \varphi{I}_{\{G_{-}<1\}}\Delta{S}>-1\}$, which follows from the condition $\{\widetilde{G}=1>G_{-}\}=\emptyset$ in (\ref{Assumptions4Tau}), we deduce that $\varphi$ is an $\mathbb{F}$-predictable and bounded process satisfying  
 $$\varphi{I}_{\{G_{-}<1\}}\Delta{S}>-1.$$
 This proves that $\varphi\in \Theta_b(I_{\{G_{-}<1\}}\is{S},\mathbb{F})$, and the proof of the lemma is complete.
\end{proof} 
\section*{Acknowledgements}
\noindent  This research is fully supported by the
Natural Sciences and Engineering Research Council of Canada, Grant NSERC RGPIN-2019-04779. \\ 
\noindent The authors would like to thank  Safa' Alsheyab, Jun Deng, and Mich\`ele Vanmaele for several comments/suggestions, fruitful discussions on the topic, and/or for providing important and useful related references.\\
\noindent The authors are very grateful to an anonymous referee for the careful reading, important suggestions, and pertinent comments that helped improving the paper. 


\end{document}